\documentclass[aps,prb,twocolumn,showpacs,floatfix]{revtex4-1}
\usepackage{graphicx}
\usepackage{color} 
\usepackage{amsmath}
\usepackage{amssymb}
\usepackage{comment}
\usepackage{setspace} 
\usepackage{tikz}
\usetikzlibrary{shapes,arrows}
\usepackage{titlesec} 
\def\nnu{{\nonumber}}
\definecolor{scarred}{rgb}{0.75,0.0,0.0}

\begin{document} 
\title{A multi-orbital iterated perturbation theory for model Hamiltonians and real material-specific calculations of correlated systems} 
\author{Nagamalleswararao Dasari}\email[E-mail:]{nagamalleswararao.d@gmail.com}
\author{Wasim Raja Mondal}
\author{N.\ S.\ Vidhyadhiraja} \email[E-mail:]{raja@jncasr.ac.in}
\affiliation{Theoretical Sciences Unit,\\Jawaharlal Nehru Centre for Advanced Scientific Research,\\Jakkur, Bangalore 560 064, India.}
\author{Peng Zhang} \author{Juana Moreno} 
\author{Mark Jarrell} 
\affiliation{Department of Physics $\&$ Astronomy and\\ Center for Computation $\&$ Technology, \\
Louisiana State University,\\ Baton Rouge, LA 70803-4001, USA.}

\begin{abstract}
Perturbative schemes utilizing a spectral moment expansion are well known and
extensively used for investigating the physics of model Hamiltonians and real material systems (in
combination with density functional theory). However, such methods are not always reliable in
various parameter regimes such as in the proximity of phase transitions or for strong couplings. 
Nevertheless, the advantages they offer, in terms of being computationally inexpensive, with
real frequency output at zero and finite temperatures, compensate for their deficiencies and offer a
quick, qualitative analysis of the system behavior. In this work, we have developed such a method, that
can be classified as a multi-orbital iterative perturbation theory (MO-IPT) to study N-fold
degenerate and non degenerate Anderson impurity models.  As applications of the solver, we have
combined the method with dynamical mean field theory to explore lattice models like the single
orbital Hubbard model, covalent band insulator and the multi-orbital Hubbard model for
density-density type interactions in different parameter regimes. The Hund's coupling effects in
case of multiple orbitals is also studied. The limitations and quality of results are gauged through
extensive comparison with data from the numerically exact continuous time quantum Monte Carlo
method (CTQMC). In the case of the single orbital Hubbard model, 
covalent band insulators and non degenerate multi-orbital Hubbard models, we obtained
an excellent agreement between the Matsubara
self-energies of MO-IPT and CTQMC. But for the degenerate multi-orbital Hubbard
model, we observe that the agreement with CTQMC results gets better as we move away from
particle-hole symmetry. We have integrated MO-IPT with density functional theory based electronic
structure methods to study real material systems. As a test case, we have studied the classic,
strongly correlated electronic material, SrVO$_3$. A comparison of density of states and photo
emission spectrum (PES) with results obtained from different impurity solvers and experiments 
yields good agreement.           
\end{abstract} 
\maketitle

\section{Introduction}

The development of efficient methods to solve quantum impurity problems, especially those
involving multiple orbitals, has been a significant research direction in the field of theoretical
condensed matter physics.
Subsequent to the development of the dynamical mean field theory (DMFT)\cite{RevModPhys.68.13}, which is
exact in the limit of infinite dimensions and an excellent local approximation in finite
dimensions, the importance of obtaining reliable solutions to general quantum impurity problems  
has increased further.

  Within the DMFT framework, a lattice model may be mapped onto a quantum impurity
embedded in a self-consistently determined host. The impurity problem may then be solved by a
variety of techniques including-- numerically exact methods like quantum Monte Carlo (QMC), numerical
renormalization group (NRG), exact diagonalization (ED) and density matrix renormalization group
(DMRG) or semi-analytical methods like iterated perturbation theory (IPT), local moment approach
(LMA), non-crossing approximation (NCA) and fluctuation exchange approximation (FLEX). Each method
has its own advantages as well as pitfalls. For example, QMC\cite{Gull} is a numerically exact
method, but is computationally expensive. It yields data on the Matsubara axis (or imaginary time) 
so to obtain dynamical quantities such as the density of states and transport quantities, analytic
continuation of the data to real frequencies is essential\cite{m_jarrell_96a}, which is a mathematically ill-posed problem. 
Additionally, it is very difficult to access the
low temperature region where statistical errors become important. As a real frequency method,
NRG\cite{RevModPhys.80.395} can avoid the difficulties that arise from the need to carry out analytic
continuation. However, the method becomes extremely cumbersome for more than one impurity or
channel. NRG is better suited for low temperature studies.
Recently,\cite{Pruschke}
NRG was applied to study degenerate multi-orbital lattice problems, but the non-degenerate case
remains unexplored. ED\cite{PhysRevLett.72.1545} is also a real-frequency method, but one considers only a finite
number of bath states, so the resulting energy spectrum is discrete, and the broadening procedure
for obtaining continuous spectra is not free of ambiguities. Moreover, large systems or
multi-orbitals are not accessible. DMRG\cite{PhysRevLett.93.246403} for the single site case has some numerical artifacts and
its accuracy as an impurity solver is not entirely clear\cite{Gull}. 

   The semi-analytical methods are perturbation theory based solvers that attempt to capture the
essential physics by constructing an ansatz for the single-particle quantities. The ansatz is based
on satisfying various limits or conservation laws, and comprises diagrams up to a certain order or
sums a specific class of processes to infinite order. The main advantages of these methods are that
they are computationally less expensive than the numerically exact methods listed above, while also
yielding real frequency data. However, semi-analytical methods are, by definition, approximate and need to be
benchmarked against exact results to gauge their range of validity. For example, although
NCA\cite{Andreas} gives qualitatively correct results for temperatures higher than the Kondo
temperature, spurious non-analyticity at the Fermi energy develops at lower temperatures\cite{PhysRevB.53.1850}. To
recover the correct Fermi liquid behavior at low temperatures, one needs to consider a larger class of
diagrams\cite{PhysRevB.64.155111}. The FLEX approximation is conserving in the Baym-Kadanoff sense, but it does not have the correct strong
coupling behavior. So when it is employed for the half-filled Hubbard model, strong coupling
physics like the Mott transition is not captured\cite{Morita2002547,RevModPhys.78.865}. The FLEX\cite{Haule} has been extended to study
degenerate multi-orbital problems but the issues plaguing single-orbital problems remain. The LMA is
a highly\cite{LMA1,LMA2} accurate technique, that has been benchmarked extensively~\cite{LMA3} with
NRG, but the method has not been used to study lattice problems except the periodic Anderson
model\cite{Pramod}. Moreover, extensions to symmetry broken phases or multiple orbital problems
remain to be carried out. 

The  IPT is a simple, second order perturbation theory based method and it
has been used widely to solve impurity\cite{Martin2,Yosida1} and lattice problems\cite{George1} at
zero as well as at finite temperature. In the IPT, a self-energy ansatz is constructed that
interpolates between known limits (i.e., weak coupling, atomic and high frequency limits) which is why
it is also called an interpolative approach. It is clear that even the single-orbital IPT is not
free of ambiguities so different constraints or limits to construct the ansatz yield
different results. Hence, an IPT for multi-orbital problems has been `synthesized' in many different
ways by various groups\cite{Kajueter2}, and we discuss these variations next.

The IPT ansatz for the self-energy $\Sigma(\omega)$
is based on a rational or continued fraction expansion of a specific
subset of diagrams, and consists of a small number of free variables that are fixed by satisfying
various limits, such as atomic and high frequency limits and conservation laws such as the Luttinger's
theorem. Such an interpolative approach  was first initiated by Mart\'in Rodero\cite{Martin1,Martin2}
for the single impurity and periodic Anderson models. The approach used the second order self-energy
as a building block and the pseudo-chemical potential $\mu_0$, was fixed by
assuming that the occupation $n_0$ of the non-interacting part of the Anderson impurity problem 
is equal to the lattice occupation $n$. Soon after the development 
of DMFT, the single band Hubbard model was studied by Muller-Hartmann\cite{muller}  
using self-consistent perturbation theory $\Sigma[G]$. The self-consistent perturbation 
method is able to produce a coherence peak in the single particle spectral function; however, it fails to 
reproduce the high energy Hubbard bands. For the single impurity
Anderson model (SIAM), Yosida and Yamada\cite{Yosida1,Yosida2} demonstrated that perturbation theory
in $U$ is quite well behaved for the symmetric case when expanding around the Hartree-Fock solution.
Based on these findings, Georges and Kotliar\cite{George1,George2} introduced an impurity solver
called iterative perturbation theory (IPT) to solve the single band Hubbard model within DMFT 
which is based on mean-field $G_0$ and  is able to capture coherent and 
incoherent features of single particle spectrum quite successfully. This is one 
of the biggest advantages of theories based on $G_0$ rather than the fully 
dressed G. Another advantage they offer is that, these theories naturally avoid 
any form of two particle divergence  and are therefore 
able to provide a reliable description also ``beyond" the perturbative 
regime\cite{PhysRevLett.110.246405}. This is very
important, because the two­particle divergencies, which might induce
ambiguities in the numerical determination of the DMFT self-energy
($\Sigma[G]$) within standard perturbation theory schemes, occur in a
rather large portion of the phase diagram, including the metallic
regime much before the Mott transition\cite{PhysRevLett.114.156402}.    


  Subsequently, Kajueter and Kotliar\cite{Kajueter1,PhysRevB.53.16214} proposed a modification to the
IPT called modified iterative perturbation theory (MIPT). In addition to the usual constraints of
IPT, the MIPT constrains the zero frequency behavior of the self energy by adding a pseudo chemical
potential $\mu_0$ to the Hartree corrected bath propagators.  This pseudo-chemical potential,
$\mu_0$, can be obtained in different ways so there is an ambiguity in the method.
Kajueter\cite{Kajueter1} fixed this free parameter by satisfying the Friedel's sum rule (equivalently
Luttinger theorem), hence his method is called IPT-L. The Luttinger theorem and Friedel's sum rule
are valid only at zero temperature, hence for finite temperature calculations,
Kajueter\cite{Kajueter2} used the same $\mu_0$ that was obtained at zero temperature. 

To study spontaneous magnetism in the single band Hubbard model, Potthoff, Wegner and
Nolting\cite{Nolting1,Nolting2} improved MIPT further by taking into account the spectral moments
up to third order and instead of fixing $\mu_0$ by using Luttinger theorem, they fixed it by the
$n=n_0$ constraint. This method may be called IPT-$n_0$. They also considered the simpler option,
where lattice chemical potential $\mu$ is equal to the pseudo chemical potential $\mu_0$.  This is
called IPT-$\mu_0$ and they bench-marked IPT-L and IPT-$n_0$ with IPT-$\mu_0$. Recently,
Arsenault, S\'emon, and Tremblay\cite{Tremblay} bench-marked IPT-$n_0$ with CTQMC and found the pathology
in IPT-$n_0$ that, in the strong coupling regime, the method does not recover a Fermi-liquid for
filling close to $n=1$. They suggested a new method (IPT-D) to fix the $\mu_0$ through a double
occupancy constraint. The range of schemes originating from the inherent ambiguities at the
single-orbital level give an idea of the far larger range of approximations that can be built at the
multi-orbital level. These schemes will be described next.

  Kajueter\cite{Kajueter2} extended his single orbital perturbative scheme to the degenerate
multi-orbital case. He used the coherent potential approximation (CPA) to calculate 
higher order correlation functions in the self energy. He showed, by benchmarking against ED, that
the scheme provides reasonable results only if the total particle density per site is less than one.
For fillings greater than one, his scheme produced a false double peaked structure at the Fermi
level instead of a single resonance. The reason for such a spurious structure is that the
high frequency tails in the continued fraction expansion can be systematically improved by
considering poles involving higher-order correlations functions in the self-energy, but this in turn
seriously degrades the low frequency behavior when the Luttinger's theorem is attempted to be
satisfied. To study quantum transport in mesoscopic systems such as multi-level quantum dots, Yeyati
et al.~\cite{Martin3} introduced an interpolative scheme based on IPT-$n_0$. Liebsch\cite{Liebsch}
applied an extension of IPT to study the orbital selective Mott-transition, using which he showed
that inter-orbital Coulomb interactions gives rise to a single first-order transition rather than a
sequence of orbital selective transitions. In   Liebsch's extension of IPT for the multi-orbital case,  he
chose the self energy to be the combination of Hartree term and second order pair-bubble diagram
with interaction vertices between electrons in different orbitals on the impurity. Laad
et al.\cite{Laad} constructed an interpolative scheme for multi-orbitals that was used extensively to
study real materials through the LDA+DMFT framework. In a similar context, Fujiwara et al.\cite{Fujiwara}
developed an interpolative approach for degenerate multi-orbitals. The novelty of their method was
that they used ligand field theory in the atomic limit to find the higher-order correlation
functions. 
  
  Although there exist a large range of schemes for extending IPT to the multi-orbital case,
extensive benchmarking of any single method has not been carried out. Recently
Savrasov et al. \cite{savrasov},  and Oudovenko et al.\cite{Oudovenko} developed an interpolative approach for
degenerate multi-orbitals based on a simple rational form of the self-energy, where the unknown
coefficients in the self-energy are determined using slave boson mean-field and Hubbard I
approximations. In their Hirsch-Fye-QMC work on the SU(4) Hubbard model, they have observed a good agreement in the particle-hole asymmetric cases.

  In the present work, we build upon the previous knowledge 
to develop an interpolative scheme for solving a general multi-orbital quantum impurity problem.  Our scheme is
also based on the second-order self-energy as a building block and we use the generic name for the
method as simply multi-orbital iterative perturbation theory (MO-IPT). Our method has a single
pseudo-chemical potential $\mu_0$, that is found by satisfying the Luttinger's theorem. We impose
the correct high frequency and atomic limits  to get the unknown coefficients in the
self-energy ansatz. In the single orbital case, we find that MO-IPT recovers the usual MIPT self
energy expression and for the degenerate multi-orbital case, our MO-IPT self-energy expression reduces to that of Kajueter\cite{Kajueter2}. The main novelty lies in handling the high frequency
poles in a systematic way. The method is general enough that it can be applied to study symmetry
broken phases, Hund's coupling (density-density type) and crystal field effects.

Since MO-IPT is a semi analytical method it needs to be bench-marked. Subsequent to the description of the method, we
embark upon an extensive benchmarking of MO-IPT with numerically exact, hybridization expansion continuous time quantum Monte Carlo method (S-CTQMC)\cite{Gull2} as implemented in the ALPS\cite{Bauer} libraries and our own implementation of interaction expansion CTQMC (W-CTQMC).  Our  main conclusion
is that the MO-IPT method works very well when used away from integer-fillings, even at reasonably
strong coupling. Using MO-IPT, we have addressed issues disputed in the current literature 
of doped Mott insulators\cite{Tremblay} and covalent band insulators\cite{werner2}. In the 
multi-orbital degenerate case, the method proposed by Kajueter\cite{Kajueter2} shows spurious 
features which restricted its applicability to fillings smaller than one. However, our approach 
circumvents all the above issues and moreover captures the filling 
dependent effect of the Hund's coupling\cite{Janus} in the low energy scale. Our study of crystal 
field effects (the non-degenerate case) is the first attempt to extend the ansatz beyond the degenerate 
case. 
In addition, we have shown that our method produces  a good default model for the analytic continuation 
of CTQMC data using the maximum entropy method~\cite{Mark3}. We have also integrated the MO-IPT with 
material-specific, density functional theory based calculations, and thus
tested it for a prototypical example of strongly correlated electronic system, SrVO$_3$. A rather good
agreement is obtained when the MO-IPT photo-emission spectra (PES) is compared with
experiments.

  We have organized the paper as follows. In section II, we outline the formalism for MO-IPT. In
section III, we discuss results  when the MO-IPT is applied in the DMFT context for lattice
problems. In section IV, we present our conclusions along with future directions and possible
improvements.
               
\section{Model and Formalism} 
\label{model} 
The multi-orbital Hubbard model for density-density type
interactions and for cubic environment in standard second quantization notation is given by
\begin{equation} 
\begin{split}
{\cal{H}}&=\sum_{i\alpha\sigma}\epsilon_{i\alpha}n_{i\alpha\sigma}+
\sum_{ij\alpha\beta\sigma}T^{\alpha\beta}_{ij}(c^{\dagger}_{i\alpha\sigma}c^{\phantom
\dagger}_{j\beta\sigma}+h.c)+\sum_{i\alpha\sigma} \frac{U}{2} n_{i\alpha\sigma}
n_{i\alpha\bar{\sigma}}\\&+\sum_{i\alpha\sigma \neq \beta\sigma'} \frac{(U-2J)}{2} n_{i\alpha\sigma}
n_{i\beta\sigma'}+\sum_{i\sigma\alpha \neq \beta} \frac{(U-3J)}{2} n_{i\alpha\sigma} n_{i\beta\sigma}\,.
\end{split} 
\label{eq:2ohm} 
\end{equation}
where $c^{\dagger}_{i\alpha\sigma}$ creates the electron at lattice site
$i$, in orbital $\alpha$ with spin $\sigma$ and $c_{j\beta\sigma}$ annihilates the electron at
site $j$, in orbital $\beta$ with spin $\sigma$.  We are mainly interested in the local single particle
electron dynamics, which is given by the momentum sum of the lattice Green's function
\begin{equation}
\hat{G}_{loc}(\omega^+) = 
\sum_{{\mathbf{k}}} \frac{1}{(\omega^+ + \mu)\mathbb{I} -
 {\hat{H}(\mathbf{k})} - {\hat{\Sigma}}(\mathbf{k},\omega^+)}\,. 
\label{eq:gloc}
\end{equation} 
Where $\omega^+$ = $\omega$+i$\eta$ and $\eta$ is the convergence factor. 
Here ${\hat{H}(\mathbf{k})}$ comprises intra-unit-cell hybridization and
inter-unit-cell hopping, namely 
\begin{align} 
{{\hat{H}}(\mathbf{k})} & = {{\hat{H}}_{intra} } +
 {{\hat{H}(\mathbf{k})}_{inter}}\\{\rm where} \hspace{0.2cm}\left({{\hat{H}}_{intra}}\right)_{\alpha\beta} & =
 \epsilon_{i\alpha} \delta_{\alpha \beta} + T^{\alpha \beta}_{ii}\;\;\\ 
 {\rm and} \hspace{0.2cm}\left({{\hat{H}(\mathbf{k})}_{inter}}\right)_{\alpha\beta} & = \epsilon(\mathbf{k})_{\alpha \beta}\,,
\end{align} 
where $\epsilon_{i\alpha}$ are orbital energies, T$^{\alpha\beta}_{ii}$ are intra-unit cell hybridization matrix elements, and $\epsilon(\mathbf{k})_{\alpha \beta}$ is the dispersion of the lattice, that
depends on its geometry. For example, in the case of a simple cubic lattice, $\epsilon(\mathbf{k})_{\alpha \beta}$ assumes the form, $-2t^{\alpha \beta}_{ij}(\cos k_x + \cos k_y + \cos k_z )$. 

  Within DMFT, one can map the multi-orbital Hubbard model on to an auxiliary impurity problem with a
self consistently determined bath. The Hamiltonian of the corresponding single impurity
multi-orbital Anderson model, is expressed in standard notation as: 
\begin{align} 
{\cal H}_{imp}=&\sum_{\alpha}(\epsilon_{\alpha}-\mu)f^{\dag}_{\alpha} f^{\phantom
\dag}_{\alpha}+\frac{1}{N}\sum_{k,\alpha} V_{k\alpha}\left(c^{\dag}_{k\alpha}f^{\phantom
\dag}_{\alpha}+f^{\dag}_{\alpha} c^{\phantom
\dag}_{k\alpha}\right)\nonumber\\&+\sum_{k,\alpha,\beta}\epsilon_{k\alpha\beta}c^{\dag}_{k\alpha}
c^{\phantom \dag}_{k\beta}+\frac{1}{2}\sum_{\alpha\neq\beta} U_{\alpha \beta} n_{\alpha} n_{\beta}
\label{eq:Hamil} 
\end{align} 
Here $\alpha$ and $\beta$ are impurity orbital indices including
spin. The first term in the above equation represents the orbital energy; the second term is the
hybridization between the impurity and the host conduction electrons, the third term represents the
host kinetic energy and the final term is the local Coulomb repulsion between electrons at the
impurity. The corresponding impurity Green's function is given by,
\begin{equation} 
\hat{G}_{imp} = \frac{1}{(\omega^+ + \mu)\mathbb{I} -\hat{\epsilon}
-{\hat{\Delta}}(\omega^+) - {\hat{\Sigma}}_{imp}(\omega^+)}\,, 
\end{equation}
where $(\hat{\epsilon})_{\alpha \beta} = \epsilon_\alpha \delta_{\alpha \beta}$. Here 
${\hat{\Delta}}(\omega^+) = 
\sum_{\mathbf{k}} |V_{k\alpha}|^2(\omega^+\mathbb{I}-{\hat{H}({\bf{k}})})^{-1}$ is the 
hybridization matrix or equivalently the self-consistently determined bath and 
${\hat{\Sigma}}_{imp}(\omega^+)$ 
is the impurity self-energy obtained by solving the impurity problem. 
The set of equations is closed by noting that, within DMFT, the
lattice self-energy is momentum-independent and is the same as the impurity self-energy, i.e
${\hat{\Sigma}}(\mathbf{k},\omega^+)={\hat{\Sigma}}_{imp}(\omega^+)$.  The local
Green's function obtained in Eq.~(\ref{eq:gloc}) is used for defining a new hybridization as
\begin{equation}
{\hat{\Delta}}(\omega^+) = (\omega^+ + \mu)\mathbb{I} - \hat{\epsilon}
- {{\hat{\Sigma}}_{imp}(\omega^+)}-{\hat{G}}_{loc}^{-1}(\omega^+)\,.  
\end{equation} 

Obtaining the self-energy however is the most
challenging step, and we employ multi-orbital iterated perturbation theory to solve the
multi-orbital Anderson model. The starting point, as usual, 
is an ansatz for the impurity self-energy, given by\cite{Kajueter1} 
\begin{align}
\left(\hat{\Sigma}_{imp}(\omega)\right)_{\alpha\beta}=\nonumber\\&\hspace{-7.5mm}\delta_{\alpha \beta}\left(\sum_{\gamma\neq\alpha} U_{\alpha \gamma} \langle
n_{\gamma}\rangle +\frac{A_{\alpha}\sum_{\gamma \neq \alpha}\Sigma^{(2)}_{\alpha
\gamma}(\omega)}{1-B_{\alpha}\sum_{\gamma\neq\alpha}\Sigma^{(2)}_{\alpha \gamma}(\omega)}\right)\,.
\label{eq:Ansat}
\end{align} 
The self-energy is thus restricted to being diagonal in the orbital basis. In the above ansatz, the first term is simply the Hartree energy and the second term contains the second order pair-bubble diagram $\hat{\Sigma}^{(2)}$ of matrix size N$\times$N,
where N is the number of orbitals. The second order pair-bubble
diagram on the real frequency axis is given by
\begin{equation} 
\begin{split}
\Sigma^{(2)}_{\alpha\beta}(\omega)= &U^2_{\alpha\beta}\int \int \int d\epsilon_1 d\epsilon_2 d\epsilon_3
\rho_{\alpha}(\epsilon_1) \rho_{\beta}(\epsilon_2) \rho_{\beta}(\epsilon_3)\\&
\frac{n_F(-\epsilon_1)n_F(\epsilon_2)n_F(-\epsilon_3)+n_F(\epsilon_1)n_F(-\epsilon_2)n_F(\epsilon_3)}{\omega^+-\epsilon_1+\epsilon_2-\epsilon_3}\,,
\end{split} 
\end{equation} 
where $\rho_{\alpha} = -\frac{1}{\pi}\mathrm{Im}\tilde{\mathcal{G}}_{\alpha \alpha}$ and
$\tilde{\mathcal{G}}_{\alpha\alpha}$ is the Hartree corrected bath
propagator, which is obtained from a Dyson like equation, and  is given by 
\begin{equation}
\mathbf{\tilde{\mathcal{G}}}^{-1}=\left(\hat{G}^{-1}_{loc}
+{\hat{\Sigma}}+\hat{\epsilon}-\left(\mu-\mu_0 \right)\mathbb{I}\right)\,.
\end{equation} 
The pseudo chemical potential, $\mu_0$, is found at $T=0$ by satisfying the Luttinger's
theorem, 
\begin{equation} 
-\frac{\rm Im}{\pi} \int^0_{-\infty} d\omega {\rm
Tr}\left(\frac{d{\hat{\Sigma}}(\omega)}{d\omega}{\hat{G}_{imp}}(\omega)\right) =0\,.
\label{eq:lutt} 
\end{equation} 
At finite temperature, an ambiguity exists in the determination of
the pseudo-chemical potential.  We choose to use the  $\mu_0$ determined at zero temperature for all
finite temperatures. The chemical potential, $\mu$, is found by fixing the total occupancy from the local Green's
function, $\hat{G}_{loc}$, to be equal to the desired filling,
\begin{equation} 
-\frac{1}{\pi}{\rm Im}\int^0_{-\infty} {\rm Tr}{\hat{G}_{loc}}= n_{tot}\,, 
\label{eq:occonst}
\end{equation}
where the trace is over the spin and orbital indices.
The unknown coefficients $A_{\alpha},B_{\alpha}$ from Eq.~(\ref{eq:Ansat}) are obtained in the
standard way by satisfying the high frequency limit and the atomic limit respectively. The
detailed procedure to derive $A_{\alpha},B_{\alpha}$ and their expressions are discussed in the Appendix.
These coefficients contain higher order correlation functions. The order of the correlation functions depends on number of poles in the self energy. For example a pole of order n involves (n+1)$^{\rm{th}}$ order correlation functions. For a two pole ansatz $A_{\alpha}$ and $B_{\alpha}$ involve two and three particle correlation functions.
We calculate the two particle correlation function~\cite{Himadri} using the equation of motion
method to obtain~\cite{Fetter}

\begin{equation} 
\sum_{m'\neq m}U_{mm'}\langle n_m n_{m'} \rangle =-\frac{1}{\pi}\int d\omega n_{F}(\omega)\mathrm{Im}\left[\Sigma_m(\omega)G_m(\omega)\right]\,.
\end{equation} 
This single equation is not sufficient to find all the two-particle correlators. Hence as an approximation,
we use the following: 

\begin{align} 
\langle n_m n_{m'} \rangle=-\frac{\int d\omega n_{F}(\omega)\mathrm{Im}(\Sigma_m(\omega)G_m(\omega))}{\pi U_{mm'}(N_{orb}-1)}. 
\end{align} 
We calculate the three particle correlation function encountered in B$_{\alpha}$ approximately by
decoupling it in terms of two and single particle correlation functions. In this work, we have ignored the three particle correlation function.

\section{Results and Discussion}
The formalism developed in the previous section is applied to a wide variety of correlated systems.  
 We begin with a discussion of the well studied paramagnetic Mott transition in the half-filled single-band Hubbard model. 
Then we examine the doped Mott insulator. The covalent insulator is considered next, followed by the two-orbital Hubbard model. 
 For the latter, We investigate the effect of filling, Hund's coupling and crystal field splitting. Finally, we move to real material 
calculations considering specifically the case of SrVO$_3$.  
As mentioned earlier we bench-mark our results with those from numerically exact
CTQMC\cite{Gull2} methods. The CTQMC formalism yields results on the Matsubara frequency axis so to
get the real frequency data, analytical continuation is required. 
We avoid analytic continuation by transforming the real frequency
data obtained from MO-IPT to imaginary frequencies using the following spectral representations: 
\begin{equation}
G(i\omega_n)=\int \frac{A_G(\omega)d\omega}{i\omega_n-\omega}, 
\label{eq:HilbertG} 
\end{equation} 
and
\begin{equation}
\Sigma(i\omega_n)=\int \frac{A_{\Sigma}(\omega)d\omega}{i\omega_n-\omega} \,,
\label{eq:HilbertS} 
\end{equation} 
where $A_G(\omega)=-{\rm Im} G(\omega)/\pi$ and
$A_{\Sigma}(\omega)=-{\rm Im} \Sigma(\omega)/\pi$.  In order to quantify the efficiency of the
method, the imaginary part of the self energy needs to be bench-marked rather than the Green's function.
This is because the former is far more sensitive than the latter and moreover, the 
low energy scale of the system depends on the imaginary part of the self energy.
     
\subsection{Single band Hubbard model: Half-filled case}
 The Hamiltonian for the single band Hubbard model is given by 
\begin{align} 
{\cal H}=&\sum_{ij\sigma}T^{\sigma}_{ij}(c^{\dagger}_{i\sigma}c^{\phantom
\dagger}_{j\sigma}+h.c)+\sum_{i\sigma}\epsilon_{i\sigma}n_{i\sigma}+\sum_{i\sigma} \frac{U}{2}
n_{i\sigma} n_{i\bar{\sigma}}\,. 
\end{align} 
We study the above model within DMFT for a semi-elliptical
density of states, given by 
\begin{equation}
\rho(\epsilon)=\frac{4}{\pi W^2} \sqrt{{\left(\frac{W}{2}\right)}^2-\epsilon^2}\,.
\end{equation} 
Here W is the full-band width. In our calculations, we choose the energy unit
to be $\frac{W}{2}$ = 1. 
\begin{figure}[htbp] 
\centering
  \includegraphics[width=\columnwidth]{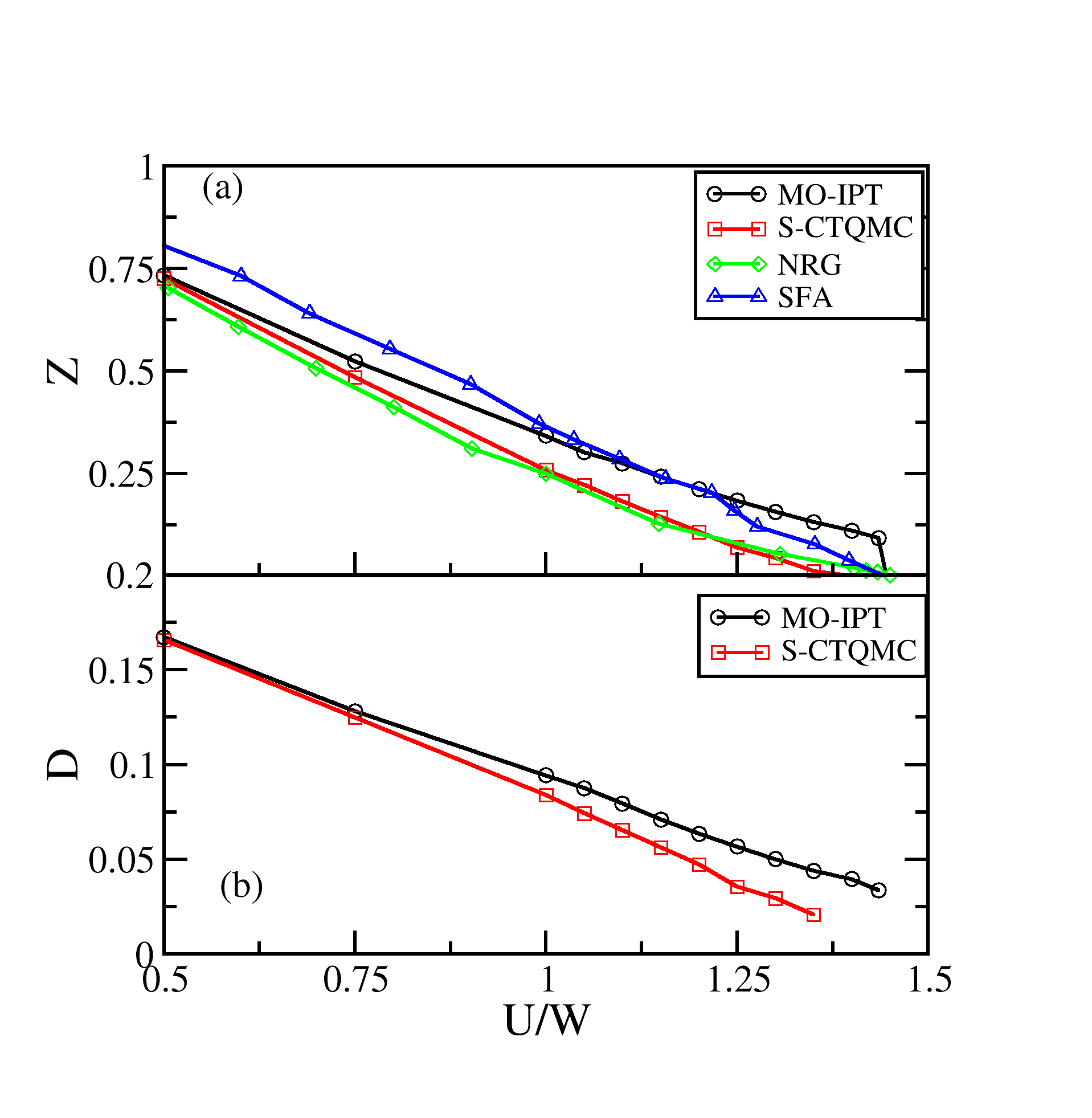}
  \caption{ (Color online) (a) Quasi-particle weight $Z$ of the single band half-filled Hubbard model 
obtained with different impurity solvers (see text for more details) 
(b) Double occupancy $D$ obtained from MO-IPT and S-CTQMC.
  \label{fig:fig1}
}
\end{figure}

\begin{figure}[htbp]
\centering 
\includegraphics[width=\columnwidth]{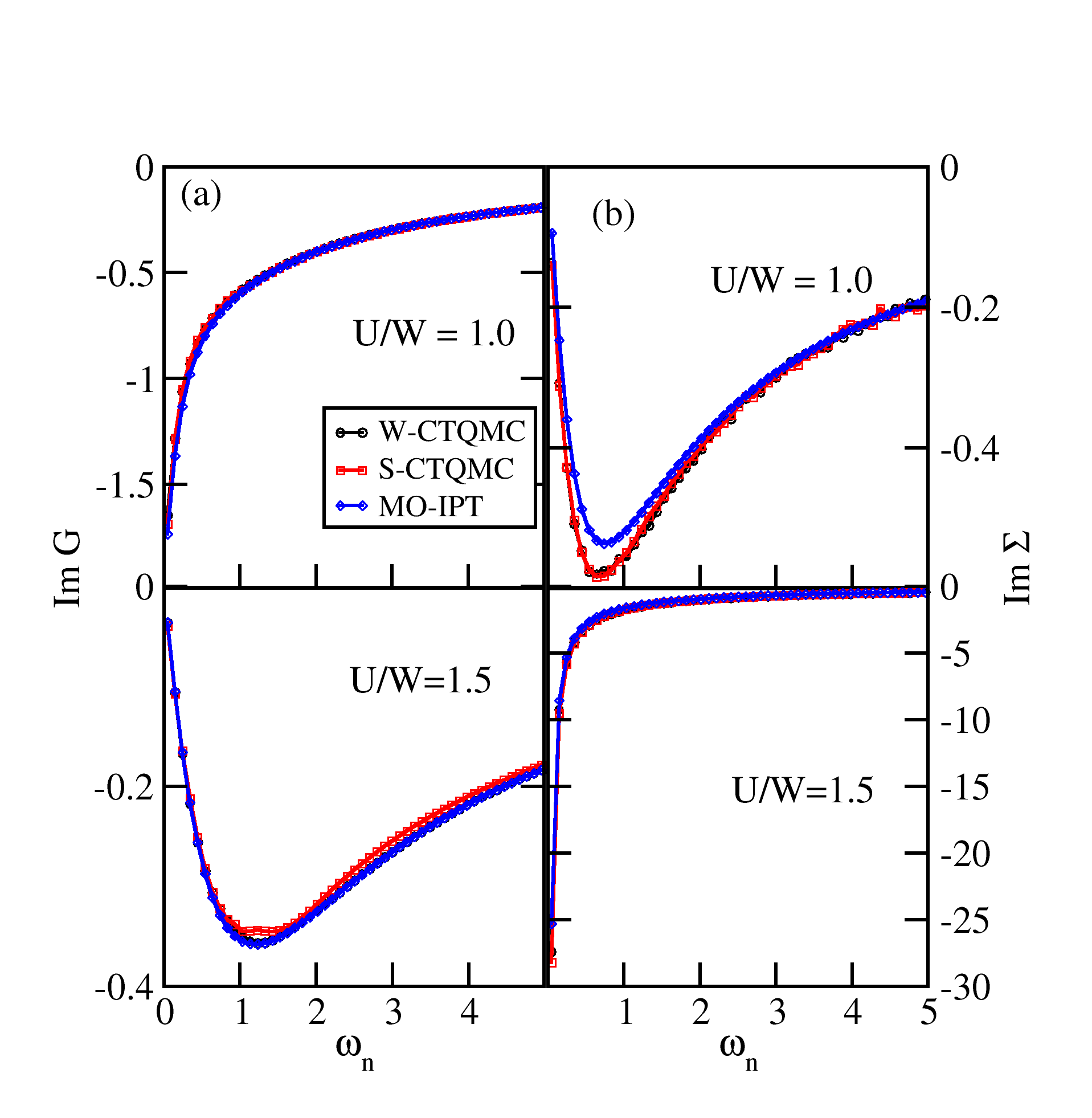} 
\caption{(color online) Comparison of the imaginary part of Matsubara Green's function (left panels) and self energy (right panels) obtained from
MO-IPT, S-CTQMC and W-CTQMC\cite{PhysRevB.87.121102} for $U/W=1.0$ (top panels) and $U/W=1.5$ (bottom panels) at $\beta$ = 64.} 
\label{fig:fig2} 
\end{figure}

  The half-filled Hubbard model exhibits an interaction-driven metal-insulator Mott transition at a
critical $U_c$. Terletska et al.~\cite{PhysRevLett.107.026401} found that the critical exponents and
scaling functions obtained by IPT are identical to those from CTQMC. Here, we revisit this case
and benchmark the quasiparticle weight, double occupancy, spectra and imaginary part of the self-energy.
The MO-IPT method reduces to the second order perturbation theory in terms of Hartree-corrected
propagators.  In Figure~\ref{fig:fig1}(a) we compare the quasi-particle weight $Z$ obtained from
different impurity solvers and several values of the Coulomb interaction. The values of $Z$ obtained
from S-CTQMC match well with those from
NRG\cite{Bulla1} for all values of U/W except close to the Mott-transition. This is most likely because
we have done CTQMC calculations at $\beta=64$, while NRG is at zero temperature. The critical
interaction strength, $\frac{U_c}{W} \approx 1.35$ obtained from both the methods\cite{Bulla2}
agrees very well. The $Z$ obtained from MO-IPT at $\beta=64$ matches quantitatively with CTQMC and
NRG in the weak coupling limit and only qualitatively in the proximity of the transition. On the
other hand, the results of the self energy functional approach (SFA)\cite{SFA} agree with MO-IPT in
the strong coupling limit rather than in the weak coupling limit. The MO-IPT yields the critical
value of $\frac{U_c}{W}$ = 1.42, which is in good agreement with the critical value $\frac{U_c}{W}$
= 1.45 obtained from SFA\cite{SFA} at zero temperature.  The double occupancy obtained from MO-IPT
and S-CTQMC (shown in panel (b) of Fig. ~\ref{fig:fig1}) also match, except very close to the
transition. A detailed comparison of spectra from S-CTQMC and W-CTQMC with
the same from MO-IPT (transformed to imaginary frequencies) is shown in Figure~\ref{fig:fig2}.
The left panels show the imaginary part of the Green's function at $U/W=1.0$ (top panel)
and $U/W=1.5$ (bottom panel), while the right panels show the imaginary part of the corresponding
self-energies. The excellent agreement between the three methods is clearly evident.

\subsection{Single band Hubbard model: Doped Mott insulator case}
 The single band Hubbard model has gained a lot of interest, because the doped Mott insulator regime is believed 
to capture the essential physics of high T$_c$ superconductors\cite{RevModPhys.78.17}. This
regime is, in reality, highly complex, because many different factors such as proximity to the
antiferromagnetic Mott insulator, disorder, d-wave superconducting fluctuations and pseudogap
physics have to be treated on an equal footing. Hence, investigations of the doped Mott insulator in
all its glory represents one of the toughest challenges in condensed matter.  Here, we take a
simplistic approach to the problem, and investigate the performance of MO-IPT in the paramagnetic
doped Mott insulator in infinite dimensions. Our MO-IPT reduces basically to the IPT-L in this
regime. 
\begin{figure}[htbp] 
\centering
\includegraphics[width=\columnwidth]{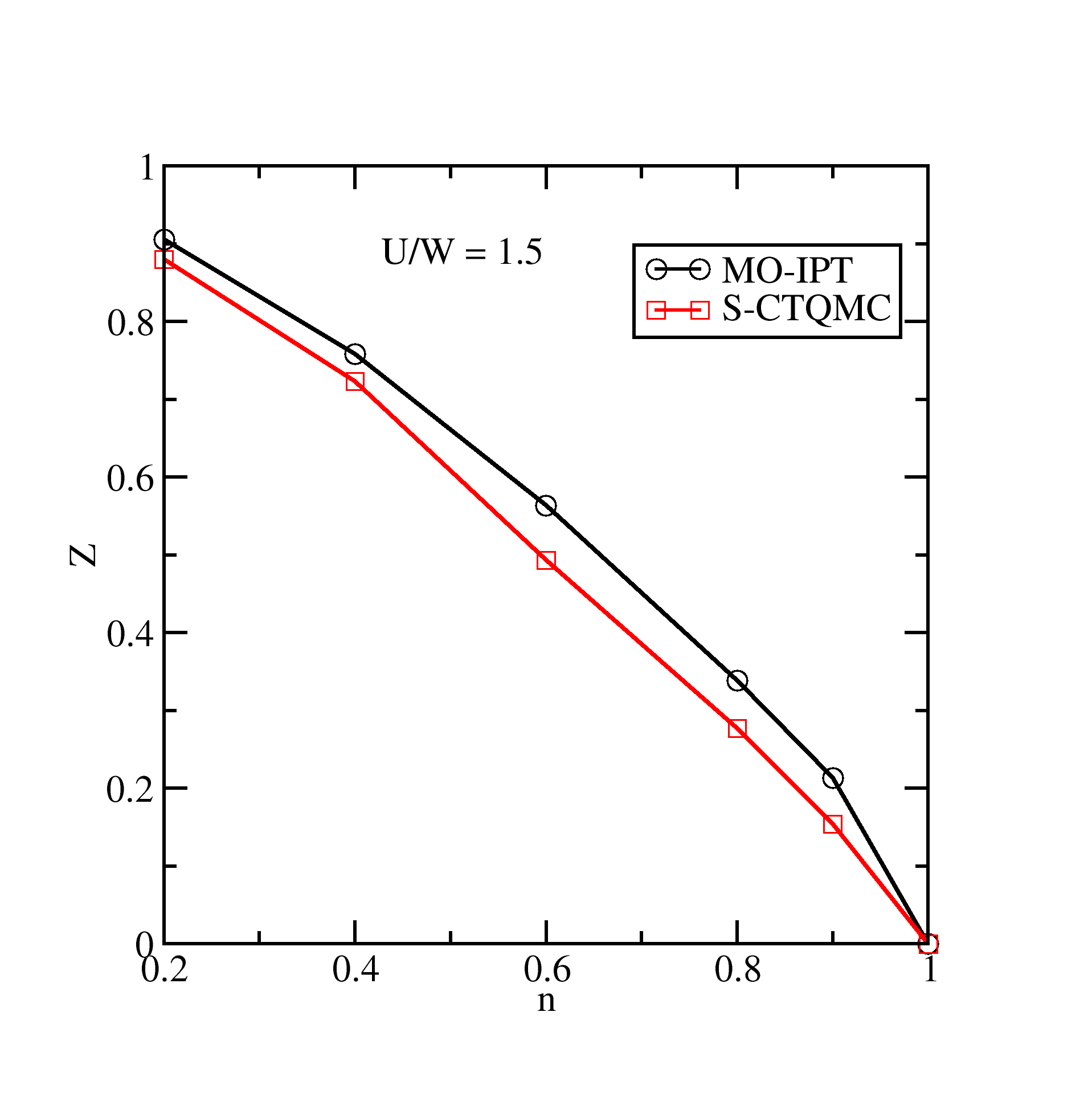}
\caption{(color online) Quasi-particle weight obtained from MO-IPT (or IPT-L) is compared to the same obtained from
CTQMC for the paramagnetic doped Mott-insulator as a function of filling with $U/W = 1.5$ and
$\beta=64$.}
\label{fig:fig3} 
\end{figure}

 A comparison of quasi-particle weight at $U/W=1.5$ obtained from MO-IPT and S-CTQMC as a function of
filling (Fig.~\ref{fig:fig3}) yields, surprisingly, an excellent agreement. We observe that as we
decrease the filling (from 1) for a given $U/W$, the Mott insulator turns into a strongly correlated
metal and finally ends up as a simple metal. In the strong coupling limit, for filling close to
$n=1$, the IPT-n$_0$ method gives an insulating solution, while the IPT-L correctly predicts a metal in agreement with exact methods. 
Kajueter and Kotliar\cite{Kajueter2} have
benchmarked the real-frequency spectral functions obtained from IPT-L with exact diagonalization
calculations and had found good agreement.
\begin{figure}[htbp] 
\centering
\includegraphics[width=\columnwidth]{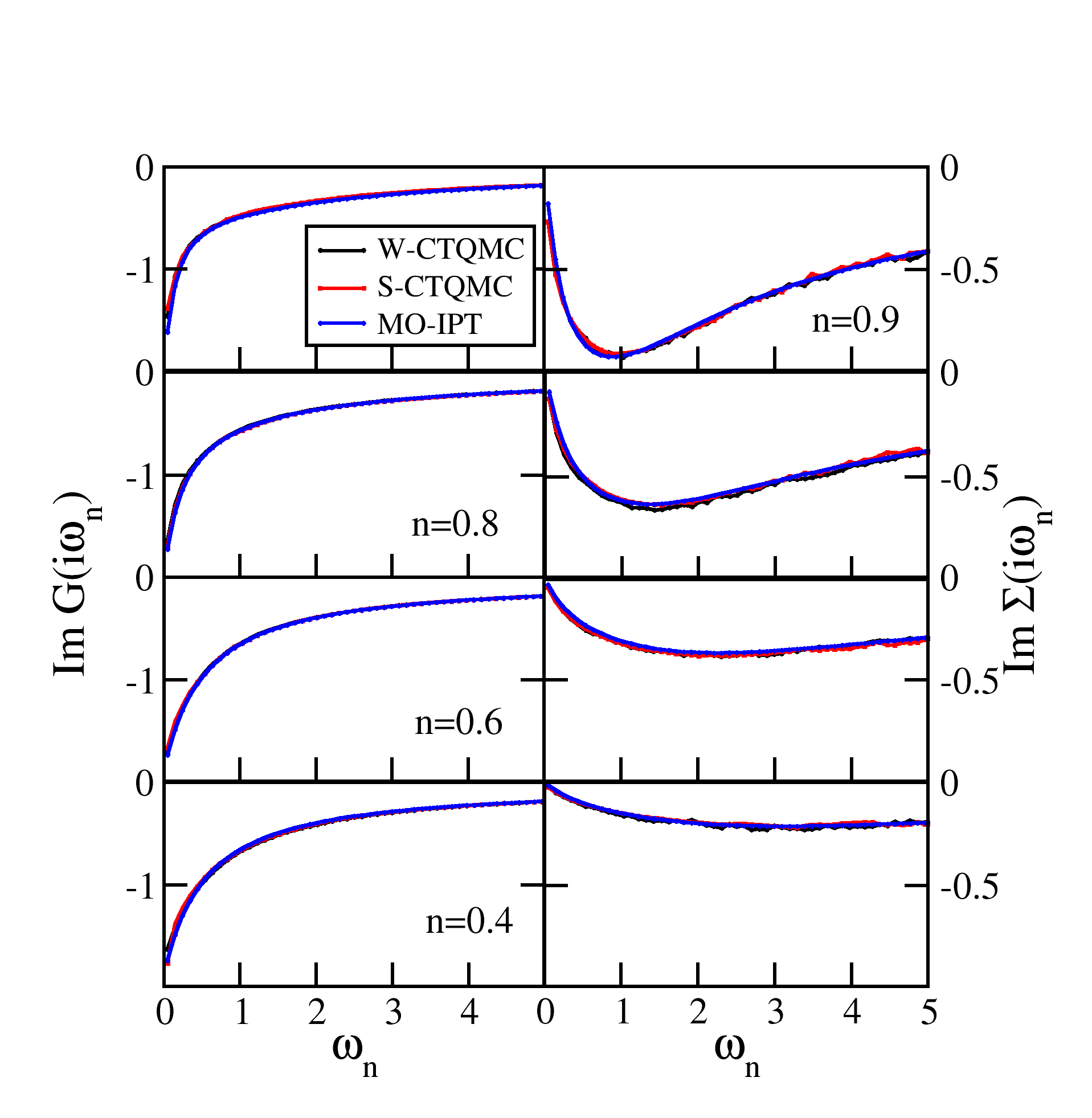}
\caption{(color online) Doped Mott insulator: Comparison of imaginary part of Matsubara Green's function and self
energy obtained from MO-IPT, W-CTQMC and S-CTQMC for U/W = 1.5 at different fillings and $\beta=64$.}
\label{fig:fig4} 
\end{figure} 
We find that the imaginary part of the Green's function and
self-energy obtained from IPT, when transformed to the Matsubara frequency axis using
Eqs.~(\ref{eq:HilbertG}), (\ref{eq:HilbertS}) are almost identical to those obtained from the strong
coupling and weak-coupling variants of CTQMC  (see Fig.~\ref{fig:fig4}). The slope of the ${\rm
Im}\Sigma(i\omega_n)$ as $\omega_n\rightarrow 0$ is $1-1/Z$, and the good agreement of $Z$ shown in
Fig.~\ref{fig:fig3} is simply a reflection of the detailed agreement for all frequencies. Such an
excellent agreement is truly surprising because IPT is a perturbative method by construction and the
strongly correlated, doped Mott insulator regime should not, in general, be amenable to perturbative
methods.


\subsection{Covalent Insulator:} 
  The discovery of topological
insulators~\cite{RevModPhys.82.3045} has led to a renewed interest in the role of e-e correlations
in band insulators (BI)\cite{kunes}.  The prime examples of such materials would be  FeSi\cite{Fesi}
and FeSb$_2$\cite{Fesb}, since experimental measurements indicate a small optical gap and large
thermopower (at low $T$). Increasing temperature leads to closing of the gap, and concomitantly a
insulator-metal crossover in the resistivity. Such large scale spectral weight transfers are highly
indicative of strong correlations. Specific heat measurements also seem to validate this
observation. The band gap in these systems is a simple consequence of the structure of the hopping
matrix and not of completely filled electronic shells~\cite{werner2}. Hence these materials are
called covalent insulators\cite{kunes,werner2}.
A Hamiltonian that describes the covalent insulator is given by\cite{werner2}
\begin{equation}
{\cal{H}} = \sum_{\mathbf{k}\sigma} \begin{pmatrix} a^{\dagger}_{\mathbf{k},\sigma} & b^{\dagger}_{\mathbf{k},\sigma}
\end{pmatrix} \hat{H}(\mathbf{k}) \begin{pmatrix} a_{\mathbf{k},\sigma} \\ b_{\mathbf{k},\sigma} \end{pmatrix} + \sum_{i\alpha}
U_{\alpha \alpha} n_{i\alpha} n_{i\alpha}\,, 
\end{equation}
where $\alpha$ = a and b are two sub-lattices with semi-elliptic bands and having dispersion 
$\epsilon_{\mathbf{k}}$  and -$\epsilon_{\mathbf{k}}$ respectively. The two sub-lattices are
coupled by a $\mathbf{k}-$independent hybridization $V$. While the unit of energy is chosen to be W = 2 throughout, for this subsection W=4 has been chosen in order to benchmark with earlier results\cite{werner2}. This is the first two-band model we have studied in
this work, since the previous cases pertained to the single-band Hubbard model. Hence this will be
the first real test of the `multi-orbital' part of MO-IPT. Since this is still the half-filled case,
Luttinger's theorem does not have to be satisfied explicitly. The $A_\alpha=1$ and  $B_\alpha =0$ for all
orbitals. Thus, the MO-IPT used for the covalent insulator case is equivalent to that employed by
Liebsch~\cite{Liebsch} for studying the Mott transition in the two-band Hubbard model. 

Before presenting the numerical results, we will present a simple analytical argument about the absence of 
any intermittent metallic phase in the CBI at zero temperature unlike the interaction-induced metallic phase in the Ionic Hubbard model\cite{PhysRevLett.97.046403}. The structure of $\mathbf{{H}_{\sigma}(k)}$ in the CBI is given by,
\begin{align}
\mathbf{{H}_{\sigma}(K)} = 
\begin{pmatrix} 
\epsilon_k & V\\
V & -\epsilon_k 
\end{pmatrix}\,,
\end{align}
and the corresponding impurity Greens function for sublattice 'a' is given by,
\begin{equation}
G_{a\sigma}(\omega^{+}) = \int d\epsilon \frac{ \zeta_{b\sigma}(\omega^{+},\epsilon)  \rho_0(\epsilon)}{\zeta_{a\sigma}(\omega^{+},\epsilon) \zeta_{b\sigma}(\omega^{+},\epsilon) - V^2}\,, 
\label{eq:green}
\end{equation}
where
\begin{align*}
&\zeta_{a\sigma}(\omega^{+},\epsilon) = \omega+ i\eta + \mu -\epsilon -\Sigma_{a\sigma}(\omega^{+})\nonumber\,, \\&
\zeta_{b\sigma}(\omega^{+},\epsilon) = \omega+ i\eta + \mu + \epsilon -\Sigma_{b\sigma}(\omega^{+})\nonumber\,,
\end{align*}
and $\eta\rightarrow 0^+$. In the half-filling case, the Hamiltonian has mirror type symmetry between sublattices, 
which reflects in the impurity Green's function and self-energy in the following way,
\begin{align}
G_{a\sigma}(\omega^{+}) & = - \left[G_{b\sigma}(-\omega^{+})\right]^{*}\,, \\
\Sigma_{a\sigma}(\omega^{+}) & = U - \left[\Sigma_{b\sigma}(-\omega^{+})\right]^{*}\,.
\end{align}
By using above self-energy symmetry relation, it is easy to see that
\begin{equation}
\zeta_{a\sigma}(\omega^{+},\epsilon) = -[\zeta_{b\sigma}(-\omega^{+},\epsilon)]^{*}\,,
\end{equation}
then Eq.~(\ref{eq:green}) can be written as,
\begin{equation}
G_{a\sigma}(\omega^{+})=\int d\epsilon \frac{\zeta^*_{a\sigma}(-\omega^{+},\epsilon) \rho_0(\epsilon)}{\zeta_{a\sigma}(\omega^{+},\epsilon) \zeta^*_{a\sigma}(-\omega^{+},\epsilon) + V^2}\,. 
\label{eq:general}
\end{equation}
With the assumption that in the band insulator and metallic phases, a Fermi-liquid expansion of self-energy holds, namely that $\Sigma(\omega)
\stackrel{\omega\rightarrow 0}{\rightarrow} \Sigma(0)
+ \omega (1-1/Z) + {\cal{O}}(\omega^2)$. Then, the value of imaginary part of self-energy at zero frequency (and zero temperature) is $
{\rm Im}\Sigma_{a\sigma}(0)=0$, and
the corresponding density of states (DOS) $D_{a\sigma}(0)=-\frac{1}{\pi} \mathrm{Im} G_{a\sigma}(0)$ is given by,
\begin{align}
&D_{a\sigma}(0)=\int \frac{d\epsilon \rho_0(\epsilon)\frac{\eta}{\pi}}{\eta^2+[\mathrm{Re}(\zeta_{a\sigma}(0,\epsilon))]^2+V^2}\,,
\label{eq:dos}
\end{align}
where $\mathrm{Re}(\zeta_{a\sigma}(0,\epsilon))$=$[\mu-\epsilon-\mathrm{Re}\Sigma_{a\sigma}(0)]$.
In the limit $\eta \rightarrow 0^+$, 
\begin{equation}
D_{a\sigma}(0)  = \int d\epsilon \rho_0(\epsilon) \delta\left(\sqrt{V^2+(\mu-\mathrm{Re}\Sigma_{a\sigma}(0)-\epsilon)^2}\right)\,.\nonumber
\end{equation}
Since the argument of the Dirac delta function is positive definite for any $V\neq 0$, the density of states at the chemical potential, $D_{a\sigma}(0)$ will necessarily vanish for any $V\neq 0$, and
hence the system will be gapped.
Thus for the CBI, interactions do not close the gap, no matter how strong they are. This implies a clear absence 
of metallicity in these insulators.
  
\begin{figure}[htbp]
\centering
\includegraphics[width=\columnwidth]{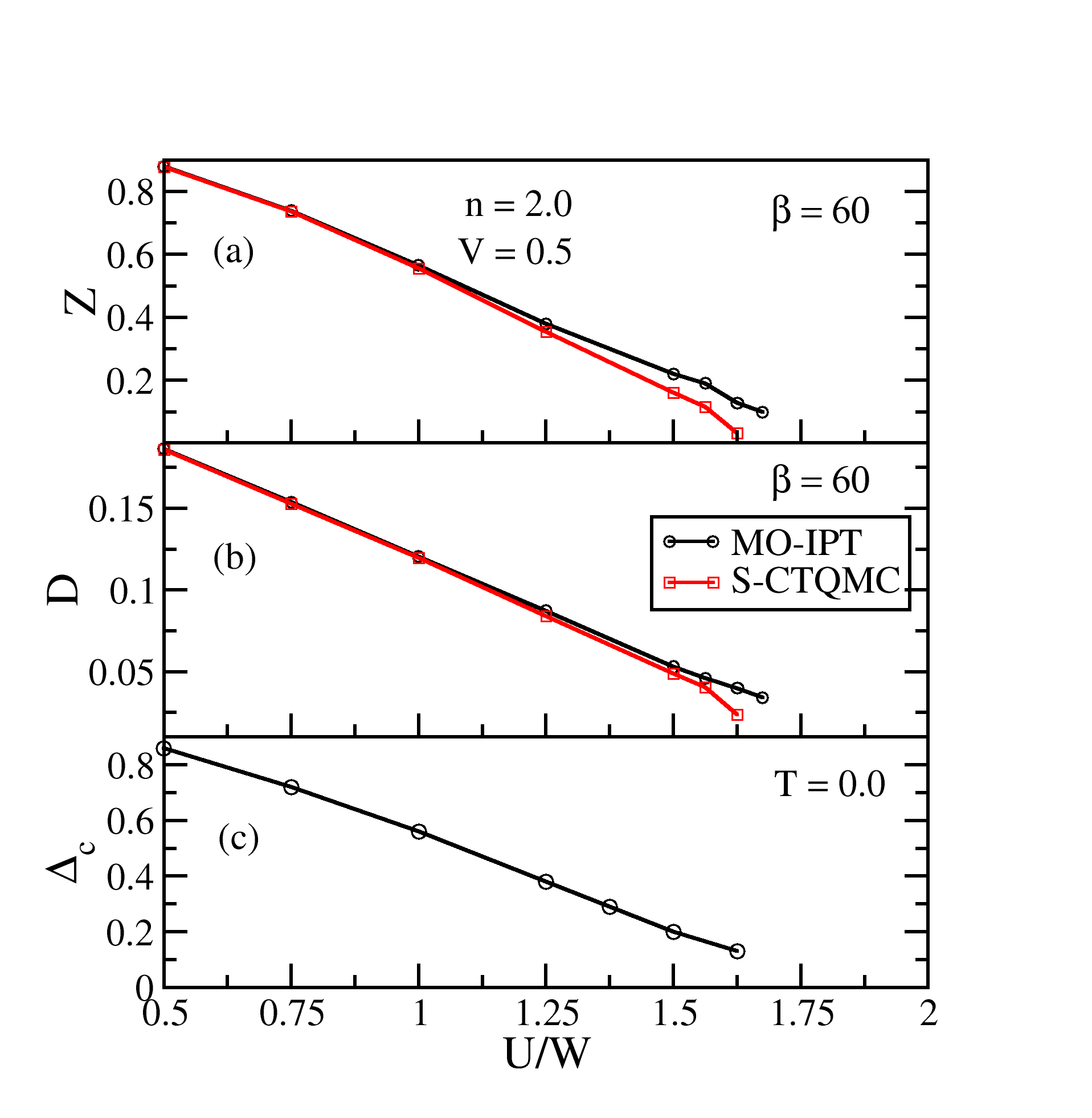}
\caption{(color online) Covalent insulator: (a) Quasi-particle weight $Z$ as a function of $U/W$ obtained from
MO-IPT (black circles) and CTQMC (red squares) for $\beta=60$ and V=0.5.(b) Double occupancy as a
function of $U/W$ obtained from MO-IPT and S-CTQMC. (c) Charge gap as a function of $U/W$ obtained
from MO-IPT at T=0.}
\label{fig:fig5}
\end{figure}

\begin{figure}[htbp]
\centering 
\includegraphics[width=\columnwidth]{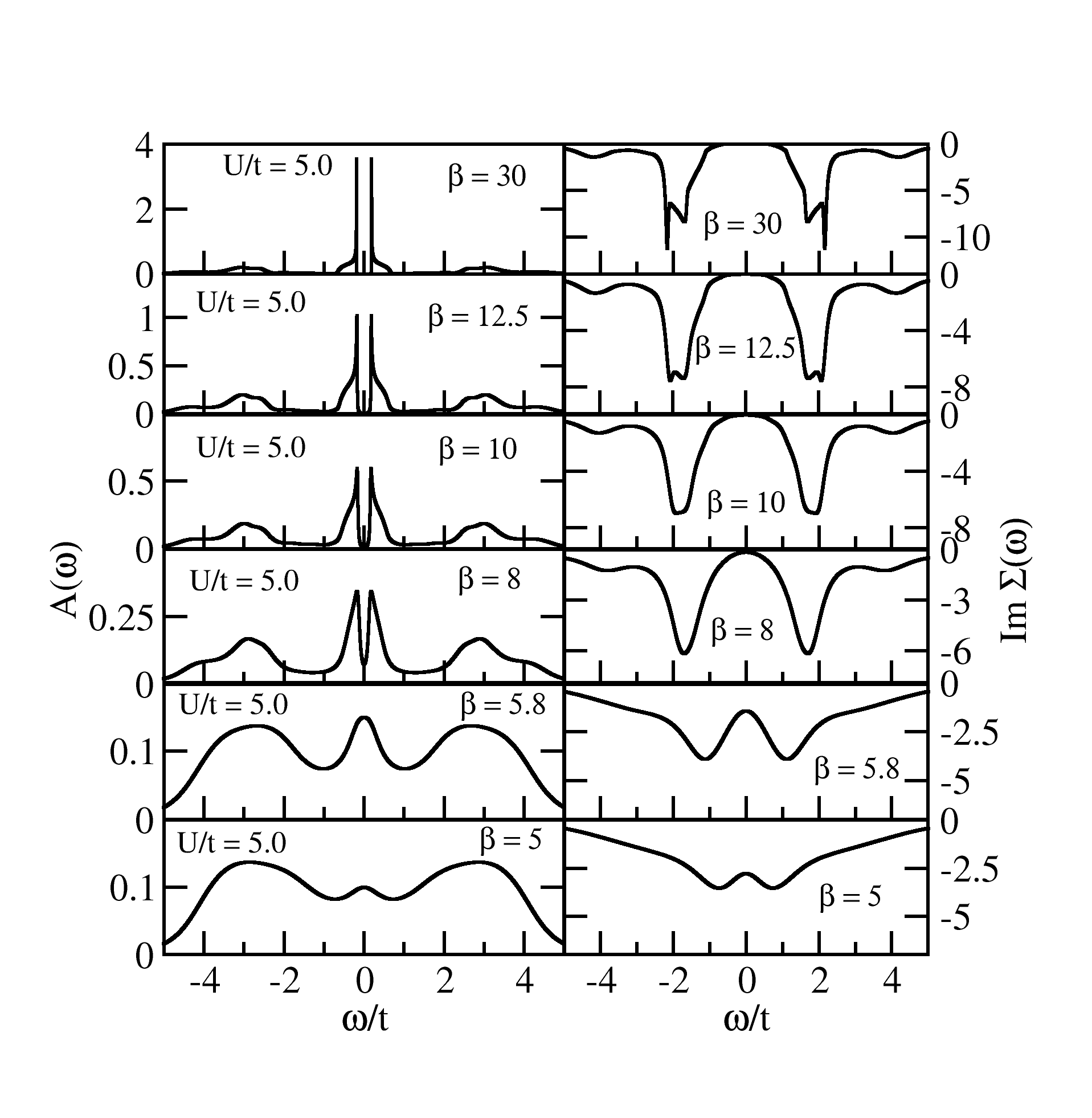} 
\caption{Covalent insulator: Spectral functions (left panels) and imaginary part of self energy of sublattice a (right panels) from
MO-IPT at $U/W = 1.25$ and V=0.5 for a range of $\beta=1/T$ values (increasing T from top to bottom). } 
\label{fig:fig6} 
\end{figure}

\begin{figure}[htbp]
\centering 
\includegraphics[width=\columnwidth]{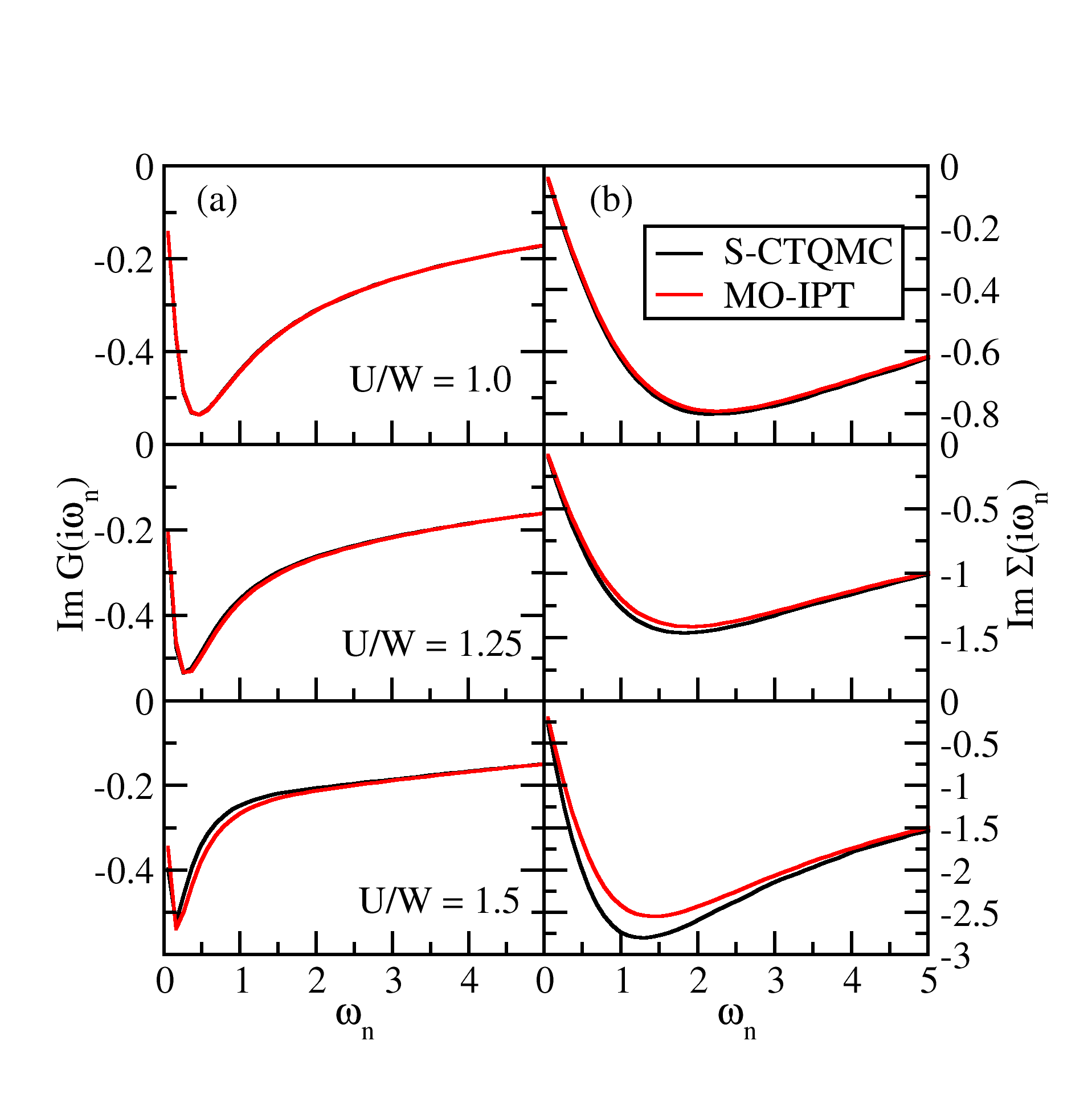} 
\caption{(color online) Covalent insulator: Comparison of the imaginary part of Matsubara (a) Green's function
and (b) self-energy of sublattice 'a' obtained from
MO-IPT (black) and S-CTQMC (red) for various $U/W$ values and $\beta$=60.} 
\label{fig:fig7}
\end{figure}

The quasi-particle weights (Fig.~\ref{fig:fig5}(a)) and double
occupancy (Fig.~\ref{fig:fig5}(b)) of sublattice 'a' obtained from MO-IPT and S-CTQMC (shown as black circles and
red squares respectively) are in close agreement except in the proximity of the transition of the
correlated band insulator to a Mott insulator. Unlike the ionic Hubbard model case\cite{PhysRevB.89.165117}, we do not see
any intervening metallic phase between the correlated band insulator and the Mott insulator. This is
also consistent with the S-CTQMC and analytical results. 
At high temperatures, the correlated band insulator should be gapless, and must develop the gap with
decreasing temperature. Precisely this behavior is seen in the real frequency spectra (left panels,
Fig.~\ref{fig:fig6}), which arises from the spectral weight transfer in the self-energy as a
function of temperature.  The high reliability of these spectra and self-energies computed through
MO-IPT is apparent in the excellent agreement with the same obtained through strong coupling CTQMC (on the
Matsubara axis, in Fig.~\ref{fig:fig7}).  The crossover of the band-insulator to Mott insulator is
also visible in the increasing (negative) slope of the imaginary part of self-energy with increasing
$U/W$. Similar kind of correlation induced transitions have also observed between 
topologically trivial and non-trivial band insulators\cite{Tsui2015330,PhysRevLett.114.185701}. 

\subsection{Two orbital Hubbard model:} 
  Encouraged by the excellent benchmarking of MO-IPT with
CTQMC for the two-band covalent insulator system, we now move on to the two-orbital Hubbard
model\cite{Antipov,Bierman}. The Hamiltonian, in standard notation, for a cubic environment and for
unbroken spin symmetry, is described in Eq. (\ref{eq:2ohm}). 
Throughout the paper, we have considered local interactions of density-density
type which are obtained by neglecting spin flip and pair-hopping terms that must be present
for a rotationally invariant Hund's coupling. The hopping is taken to be diagonal in orbital indices for simplicity.  

\subsubsection*{\bf{ (a) Half-filling: J = 0}} 
 We begin by considering the half-filled case (total
occupancy is two) with $J=0$. The Hamiltonian (Eq.~(\ref{eq:2ohm})) has SU(4) symmetry in this
situation.  We have employed a semi-elliptic non-interacting density of states of full-band width
W = 2 for the MO-IPT-DMFT calculations.
\begin{figure}[htbp]
\centering
\includegraphics[width=\columnwidth]{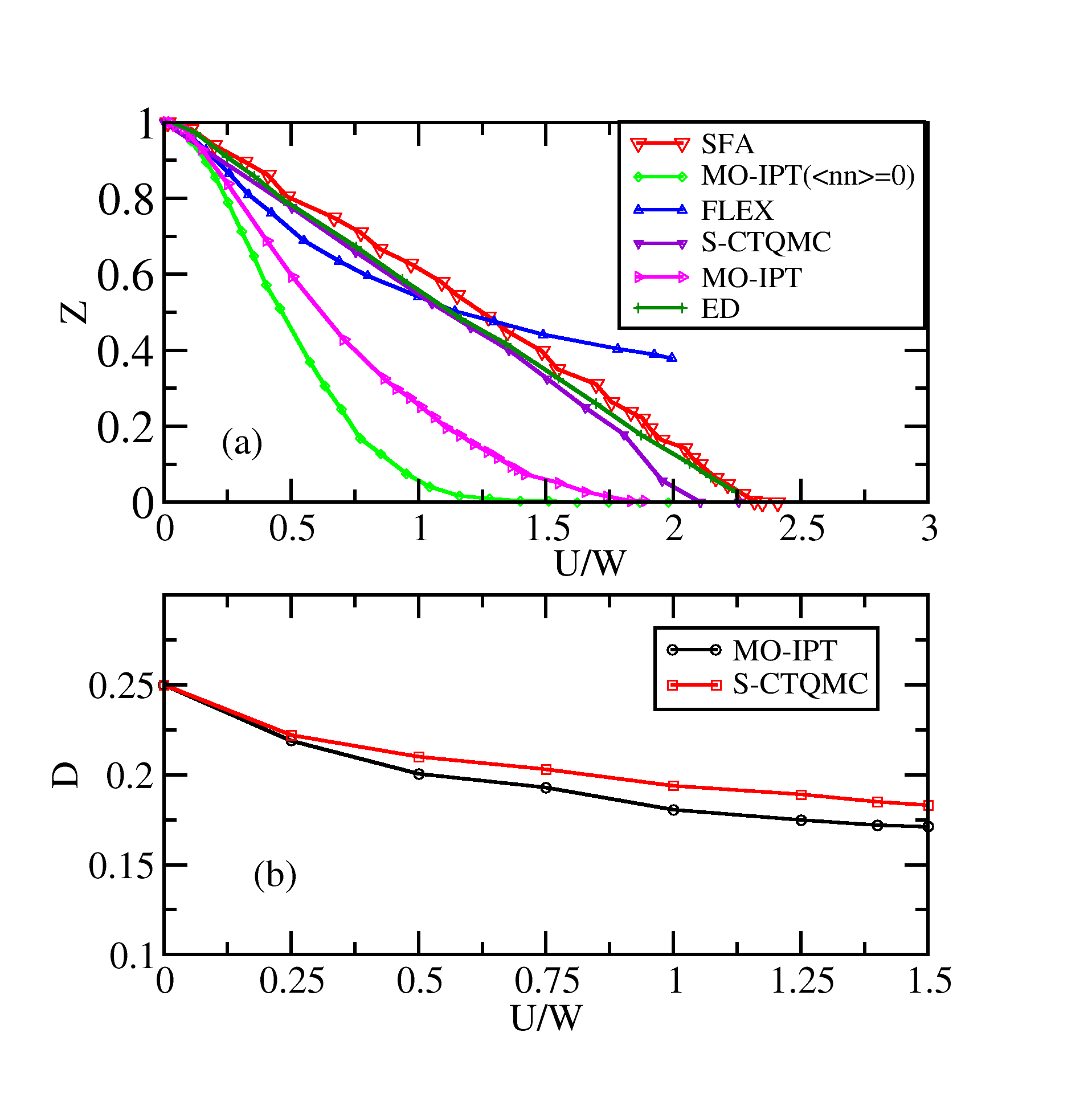}
\caption{(color online) (a) Two-orbital SU(4) symmetric Hubbard model at half-filling: 
Quasi particle weight as a function of $U/W$ obtained from different impurity solvers(SFA,MO-IPT,FLEX,S-CTQMC at $\beta=64$ and ED at T=0.0).
(b) Double occupancy obtained from MO-IPT (black circles) and hybridization 
expansion CTQMC (red squares) for $\beta$=64.} 
\label{fig:fig8} 
\end{figure}

\begin{figure}[htbp]
\centering
\includegraphics[width=\columnwidth]{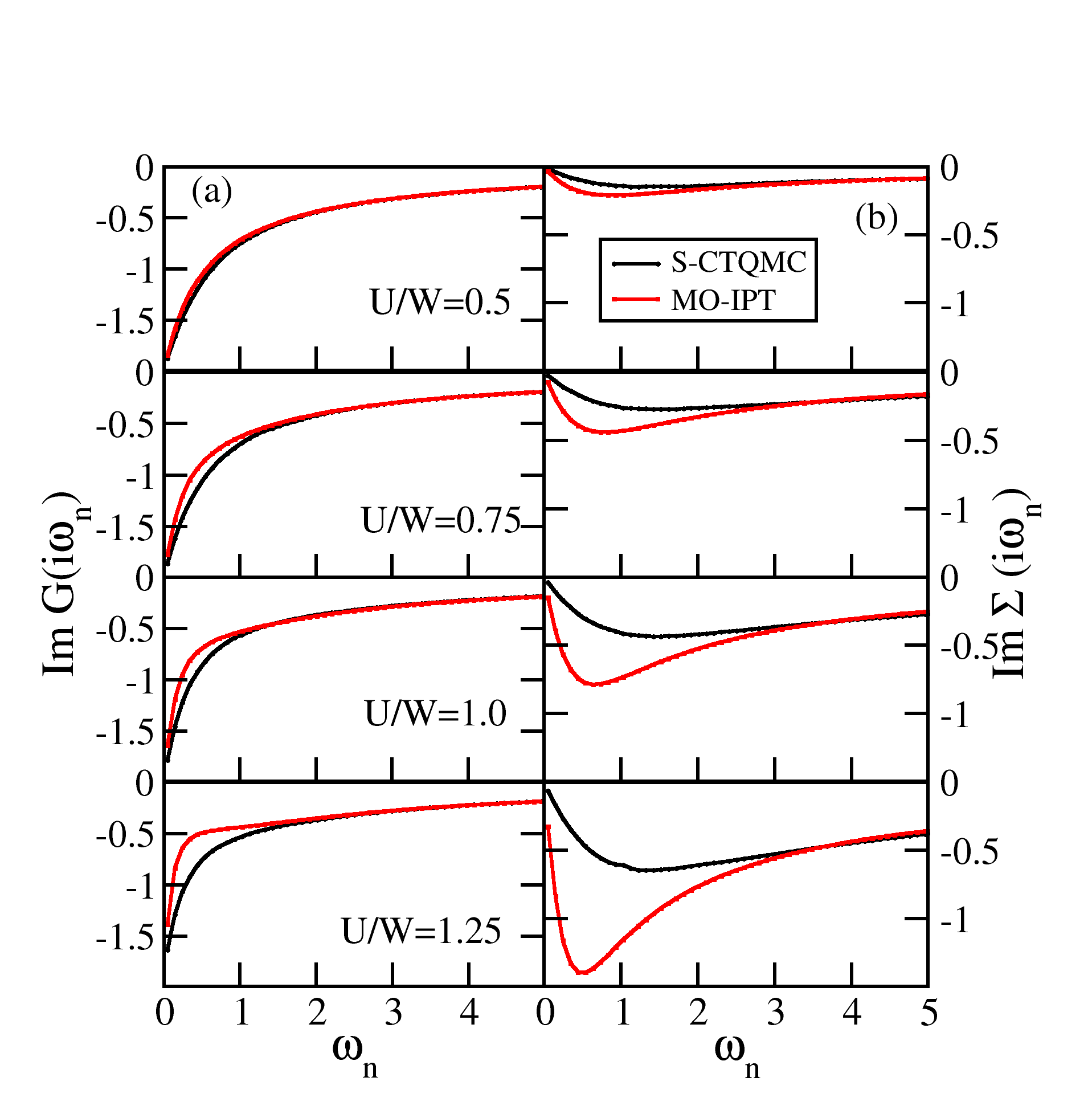} 
\caption{(color online) Two-orbital SU(4) symmetric Hubbard model at half-filling: 
Imaginary part of Matsubara Green's function (left panels) and self energy (right panels) 
obtained from MO-IPT (red
solid lines) and S-CTQMC (black solid lines) at $\beta$=64.} 
\label{fig:fig9} 
\end{figure}

 In Fig.~\ref{fig:fig8}a, we plot the quasi-particle weight ($Z$) obtained from different
impurity solvers for the particle-hole symmetric case. The results from strong coupling CTQMC, 
ED\cite{Medici} and
SFA\cite{SFA}, including the critical U$_c$, where the system transitions from metal to
Mott-insulator, are in good agreement. The critical value U$_c$ obtained in the multi-orbital case
is greater than the value obtained in the single band case. The Mott transition is absent in the
FLEX result~\cite{Haule}.
The MO-IPT is seen to underestimate the quasiparticle weight as compared to the other methods
(except MO-IPT$<$nn$>$=0; see below). However, the critical $U_c$ agrees reasonably well with that from hybridization expansion CTQMC. The green
diamonds are from a variant of MO-IPT (used e.g. by Fujiwara et al.~\cite{Fujiwara}) where the two-particle
correlation function is simply decoupled into two single-particle terms ($\langle n_\alpha n_\beta
\rangle$=$\langle n_\alpha \rangle \langle n_\beta \rangle $).  The neglect of two particle
correlations leads to a much worse comparison than MO-IPT. 
In contrast to the not-so-good agreement with exact methods for the quasiparticle weight, the
average double occupancy obtained from MO-IPT shows excellent agreement with CTQMC (see
Fig.~\ref{fig:fig8}b). Since the total energy of the system depends on single particle and two
particle correlation functions, we expect that thermodynamic quantities like total energy or
specific heat computed through MO-IPT might be reliable. One more important observation is that
the double occupancy remains finite and almost constant even beyond the Mott transition, unlike the
the single band case. We also compare the single-particle Green's function and self-energy on the
Matsubara frequency axis (Fig.~\ref{fig:fig9}). At high frequencies, the agreement between MO-IPT
and S-CTQMC is seen to be excellent, while the agreement worsens at low frequencies, especially with
increasing $U/W$.  

\subsubsection*{\bf{(b) Half-filling: Effect of Hund's coupling ($J$)}}
 The interplay of Hund's coupling, $J$, and local interaction, $U$, has been investigated by several
groups. The main consensus is that strong correlation effects can be affected significantly through
$J$~\cite{Medici,Janus}.  For example, in the half-filled case, the $U_c$ for Mott transition is
lowered by $(N-1)J$, where $N$ is the number of orbitals, while the critical $U$ is enhanced by $3J$
in the non-half-filled (but integral occupancy) case~\cite{Medici}.  
\begin{figure}[h]
\centering 
\includegraphics[width=\columnwidth]{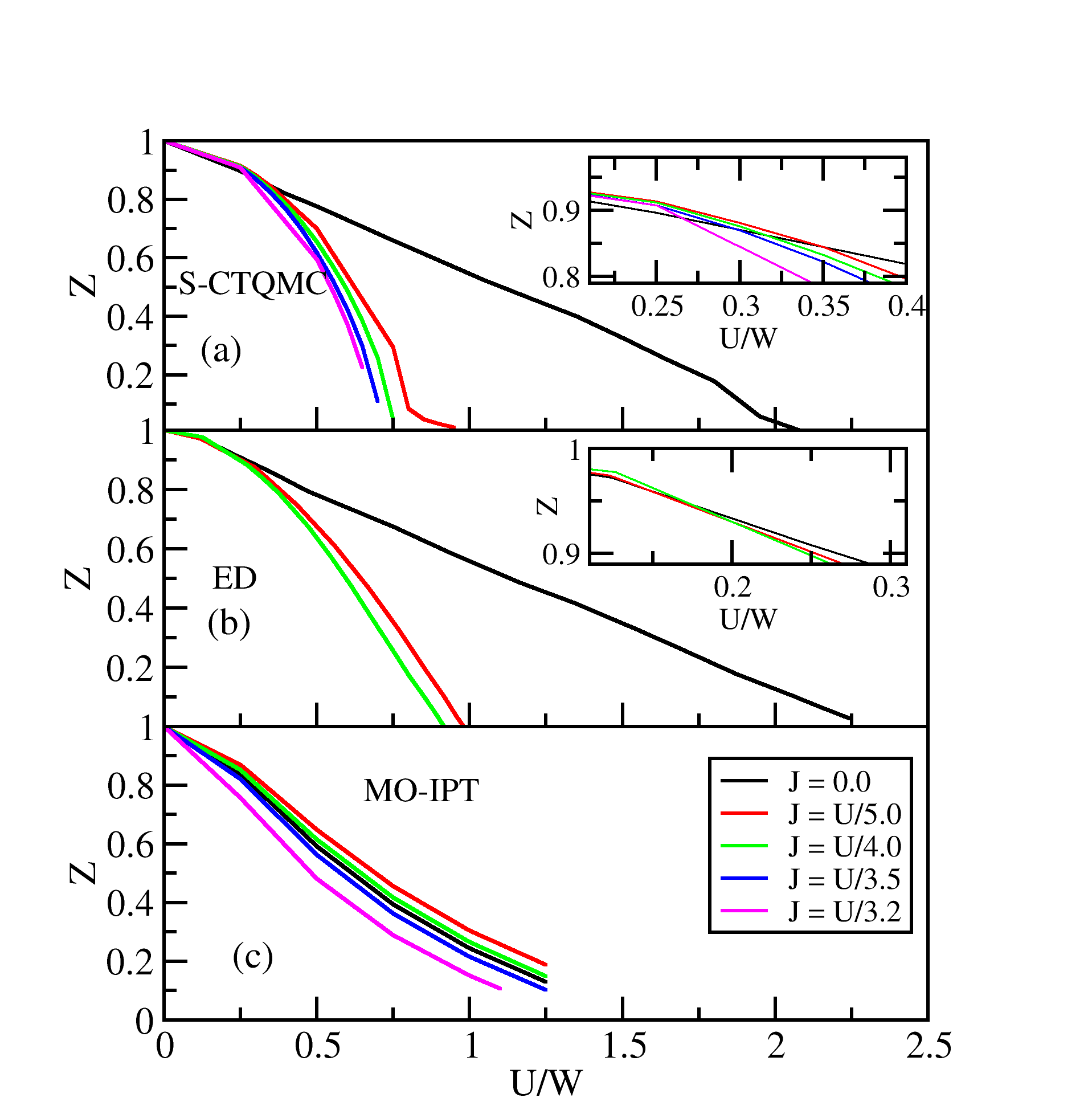}
\caption{(color online) Two orbital half-filled Hubbard model, finite $J$.
Quasi particle weight dependence on $U/W$ obtained from (a) strong coupling CTQMC, 
(b) ED  and (c) MO-IPT for various $J$ values. Insets in the panels
(a) and (b) show the effect of $J$ on $Z$ in the weak coupling regime.} 
\label{fig:fig10} 
\end{figure} 
It is important to know the extent to which the interplay between $J$ and $U$ is captured by the MO-IPT
method.  In Fig.~\ref{fig:fig10}(a) and (b), the quasi-particle weight $Z$ for different values of J
obtained from S-CTQMC and ED~\cite{Medici} is shown.  Indeed, with increasing $J$, the $U_c$ at
which $Z\rightarrow 0$ decreases sharply, as expected from the atomic limit. Also, for each $J$, the
quasiparticle weight decreases monotonically with increasing interaction strength. Although the
latter trend is qualitatively captured by the MO-IPT result (shown in Fig.~\ref{fig:fig10}c) for
larger $J$, there is a disagreement with the exact results at lower $J$ values.  The MO-IPT yields a
$U_c$ that is a non-monotonic function of $J$. 
The insets of panels a and b in  Fig.~\ref{fig:fig10} zoom in on the low interaction
($U/W\lesssim 0.3$) part of the main panels.  Unlike  for $U/W\gtrsim 0.3$, where increasing $J$
leads to a monotonic reduction of $Z$, a rise and fall of $Z$ is observed for  $U/W\lesssim 0.3$.
Although such a trend is achieved by MO-IPT as well, the non-monotonicity sustains even for larger
$U/W$.  A frozen local-moment phase is seen in the S-CTQMC calculations for any given $J$ in the
strong coupling limit, while such a phase is not observed either by ED\cite{Medici} or MO-IPT calculations. It
must be mentioned here that the CTQMC calculations employ a density-density type Hund's coupling,
while the ED employs a fully rotationally invariant $J$. 
Although the quasiparticle weight dependence on $U$ and $J$ is not accurately captured by MO-IPT,
the single-particle dynamics on all scales is in qualitative agreement with S-CTQMC calculations (as seen
in Fig.~\ref{fig:fig12}). 
\begin{figure}[htbp]
\centering
\includegraphics[width=\columnwidth]{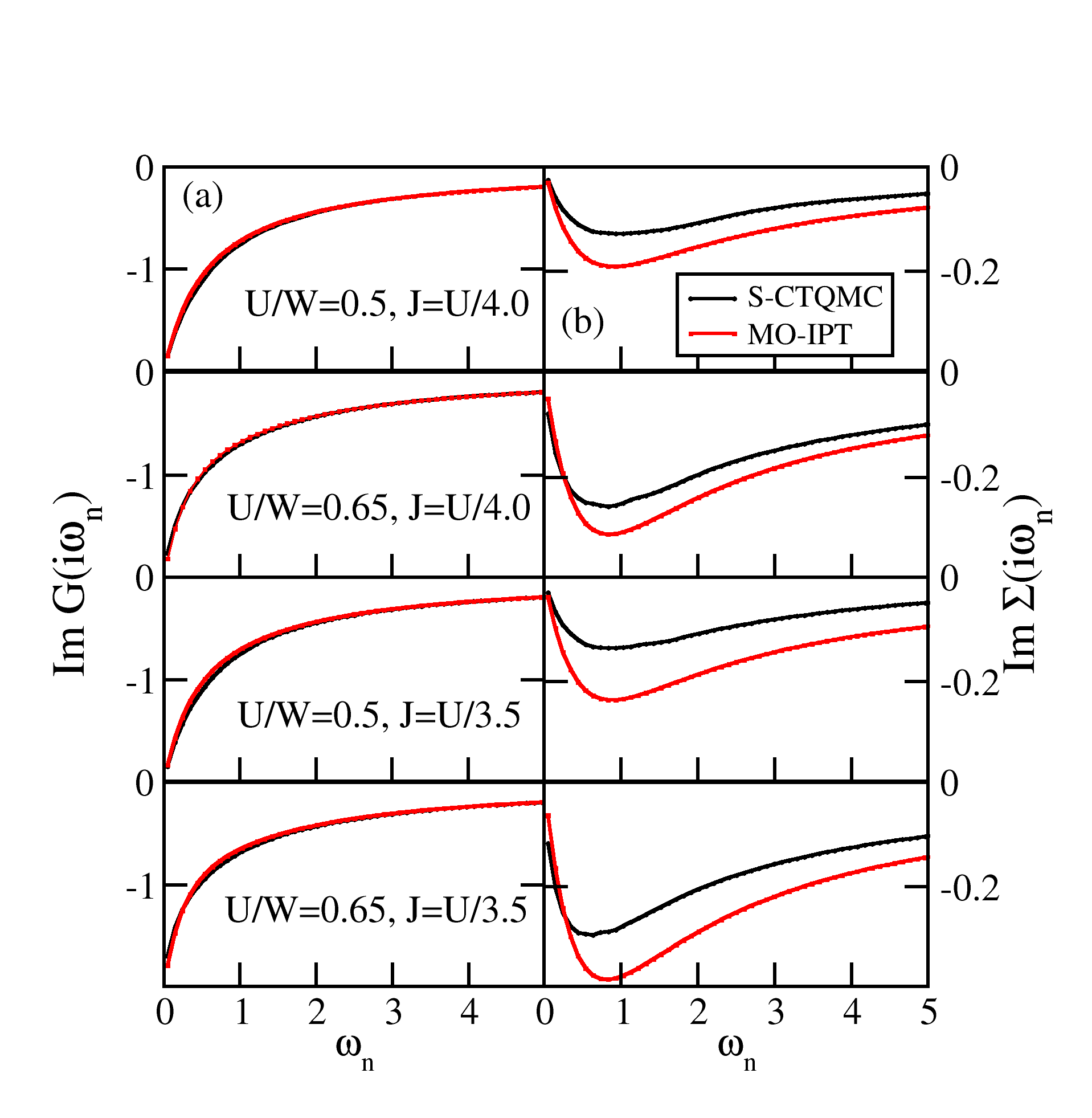}
\caption{(color online) Two orbital half-filled Hubbard model, finite $J$.
Imaginary part of Matsubara Green's
functions (left panels) and self-energy (right panels) obtained from S-CTQMC (black) and MO-IPT
(red) for different values of $J$ and $U/W$ at $\beta$=64.} 
\label{fig:fig12} 
\end{figure}
\subsubsection*{\bf{(c) Away from Half-filling: Effect of $J$ }}

\begin{figure}[htbp]
\centering
\includegraphics[width=\columnwidth]{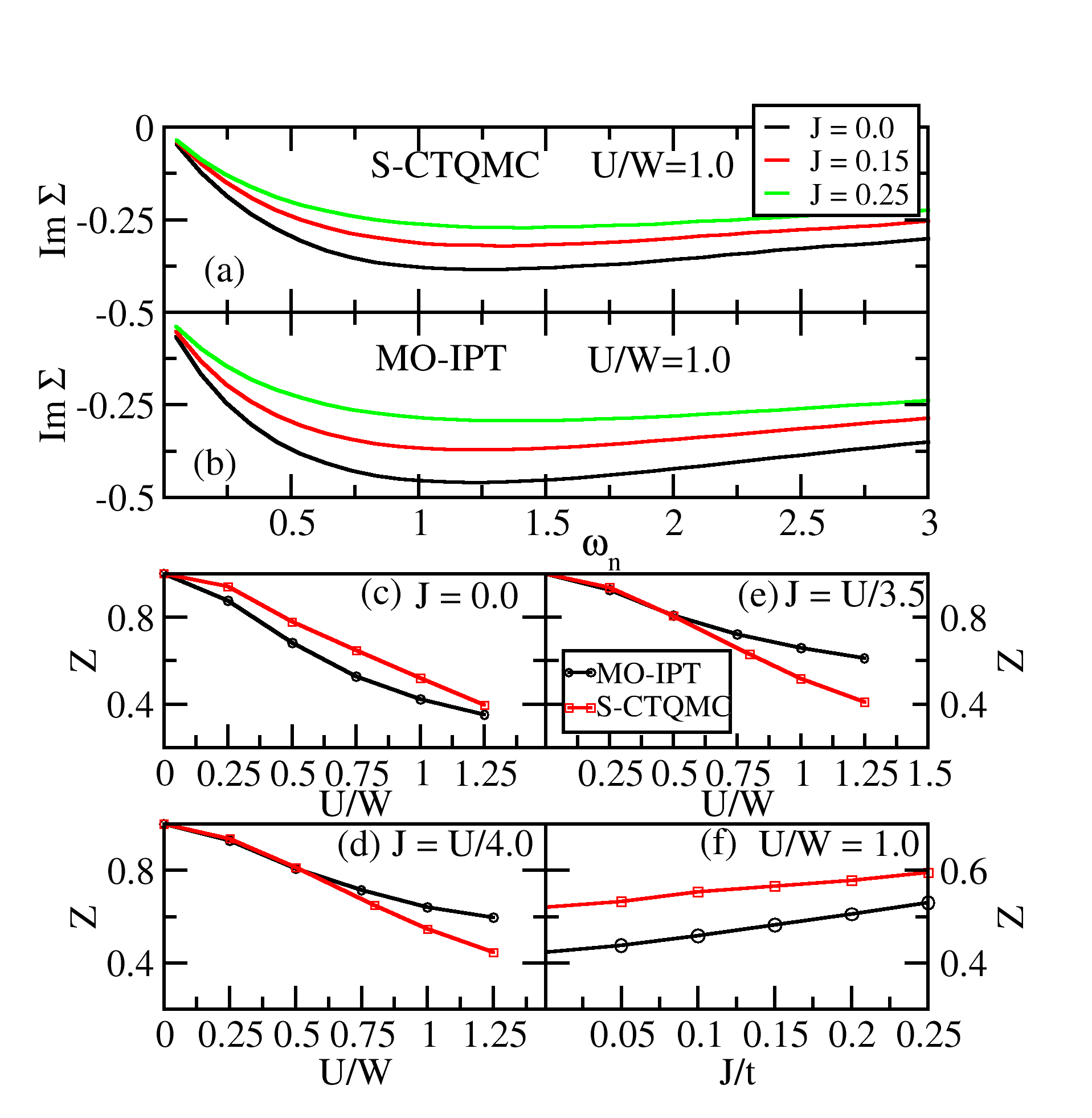} 
\caption{(color online) Two-orbital Hubbard model: Effect of $J$ away from half-filling ($n_{tot}=1.1$). The
imaginary part of the Matsubara self-energy for various J-values, and fixed $U/W=1$ as computed
within (a) S-CTQMC and (b) MO-IPT. Comparison of quasi particle weight obtained from MO-IPT (black
circles) and CTQMC (red squares) as a function of $U/W$ for (c) $J = 0.0$, (d) $J=U/4$ and (e)
$J=U/3.5$ for $\beta=64$; and (f) as a function of $J$ for a fixed $U/W=1.0$. } 
\label{fig:fig13} 
\end{figure}

  The MO-IPT method works best away from half-filling, which is consistent with the results of
comparisons carried out previously by other groups\cite{kotliar}. In order to illustrate this, here
we study the two orbital Hubbard model for a total occupation of $n_{tot}=1.1$.  The imaginary part 
of the Matsubara
self-energy obtained from S-CTQMC matches well with that from MO-IPT (Fig.~\ref{fig:fig13}, panels
(a) and (b)), hence the latter does well in this regime. This observation is reinforced by the
panels (c)-(e), which show a comparison of the quasiparticle weights as a function of $U/W$ for
three values of $J$, namely $J=0, U/4.0$ and $U/3.5$.  The results of MO-IPT are seen to agree very
well with those from CTQMC.  For most real material calculations, the regime considered in this
subsection is perhaps the most relevant. Hence, accurate results from MO-IPT  prove
its efficacy for integration into first-principles approaches.
The Hund's coupling and Coulomb interaction have a synergistic effect at half-filling, while in the
doped case, the reverse occurs\cite{Janus}. This is shown in panel (f) of Fig.~\ref{fig:fig13},
where an increase of $Z$ is seen with increasing Hund's coupling at a fixed interaction strength. 
\begin{figure}[htbp]
\centering 
\includegraphics[width=\columnwidth]{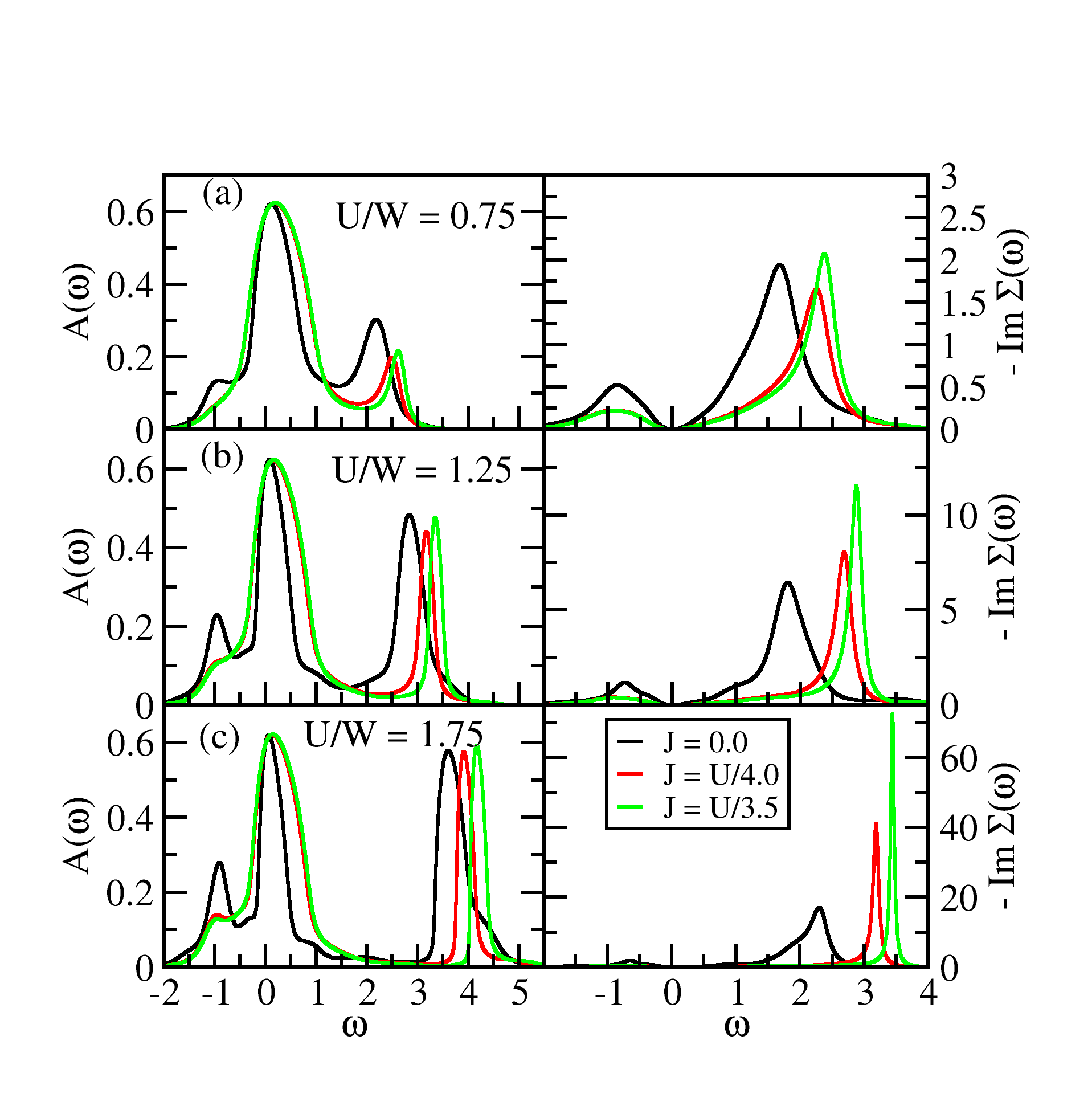}
\caption{(color online) Two-orbital Hubbard model away from half-filling. Real frequency spectral functions (left panels) and minus imaginary part of self energy (right panels) for various $U/W$ and J values.}
\label{fig:fig14} 
\end{figure} 
It is quite instructive to study the real frequency spectral
functions and self-energies as obtained from MO-IPT. These are shown in Fig.~\ref{fig:fig14} for
various values of interaction strength and Hund's coupling, $J$.  In the absence of Hund's coupling,
the spectrum (shown in the left panels of Fig.~\ref{fig:fig14}) exhibits spectral weight transfers
characteristic of increasing correlation strength: a central resonance that becomes sharper, and
Hubbard bands that grow in prominence with increasing $U/W$. However, at a fixed $U/W$, increasing
Hund's coupling leads to a reversal of the aforementioned trend, i.e, a broadening of the resonance
and a melting of the Hubbard band (see e.g. left panel bottom figure). In this parameter regime, a previous
formulation of the multi-orbital iterated perturbation theory\cite{kotliar} found a double peak
structure at the chemical potential. Such a feature was shown by the authors\cite{Kajueter1} to be
spurious by comparison to results from exact diagonalization. The reason we do not observe such a
spurious feature is that we have considered only two poles in the self-energy, in contrast to the
formulation of Ref.~\onlinecite{Kajueter1}, where they have retained all the eight poles (for a two-orbital
model). Although our ansatz seems like an ad-hoc truncation scheme, the justification for such a
scheme lies in its excellent agreement with CTQMC results (shown in Fig.~\ref{fig:fig15}) and the
absence of spurious features.  In Fig.~\ref{fig:fig15}, the imaginary part of Matsubara Green's
functions and self energies obtained from MO-IPT are compared with those from CTQMC for three
values of $J$ at $U/W=1.25$ and $\beta=$64. For all values of the Hund's coupling, an excellent
agreement is obtained. 
\begin{figure}[htbp] 
\centering
\includegraphics[width=\columnwidth]{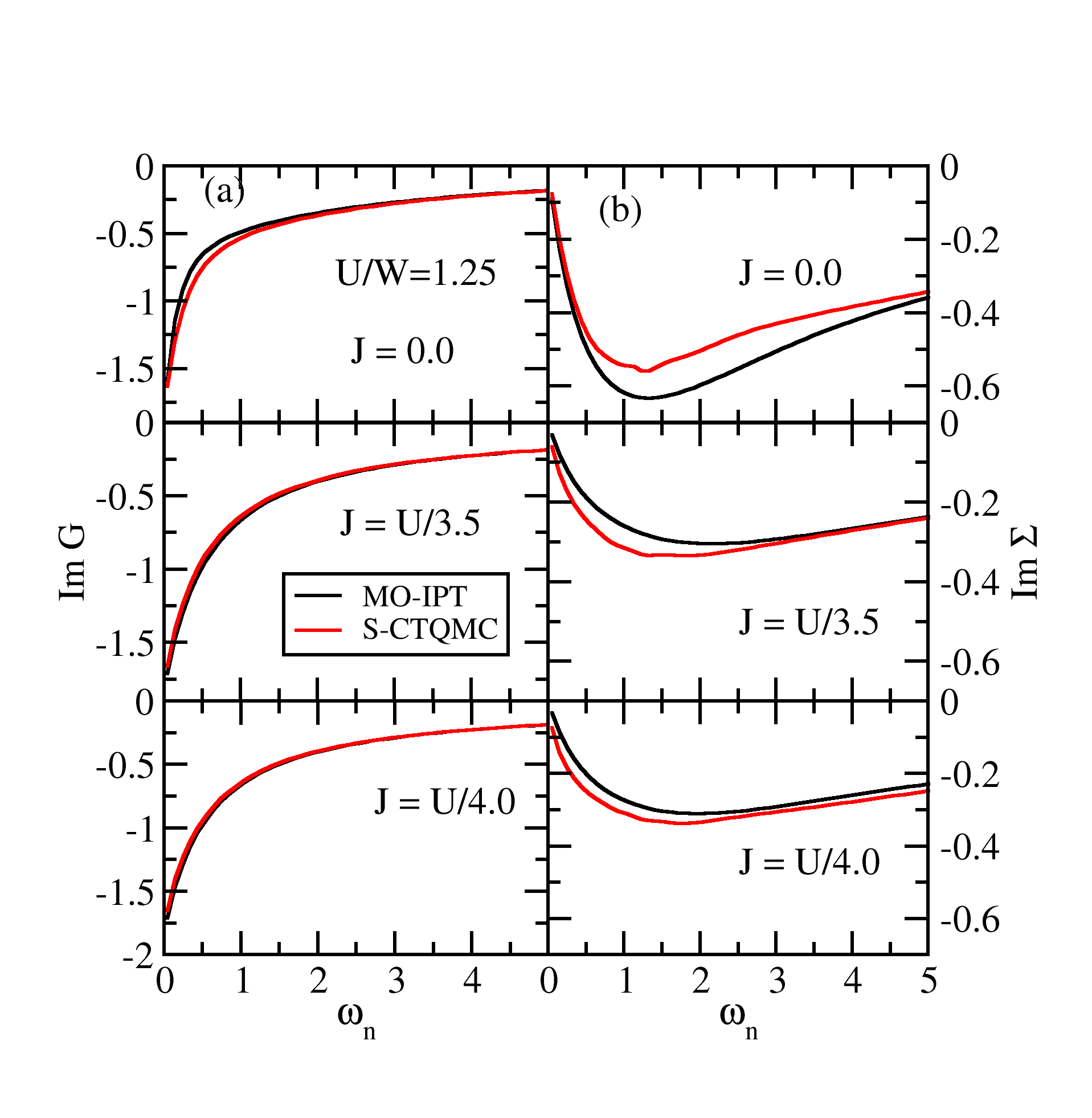}
\caption{(color online) Two-orbital degenerate Hubbard model away from half-filling ($n_{tot}=1.1$). 
Comparison of the imaginary part of the Matsubara Green's function (left panels) and the self energy 
(right panels) obtained from MO-IPT and S-CTQMC for various values of $J$ at $U/W=1.25$.} 
\label{fig:fig15} 
\end{figure}

\subsection{Two orbital Hubbard model: Crystal field splitting and Hund's coupling}

 We now proceed to the case of a non-degenerate two-orbital model with crystal
field splitting\cite{Nondeg} in the presence of Hund's coupling.  In most materials, the crystalline
environment lifts the orbital degeneracy\cite{pavarini1}. For example in transition metal oxides, due to
crystal field effects, the five fold degenerate $d$-level splits into triply degenerate t$_{2g}$ and
doubly degenerate e$_{g}$ levels and the corresponding energy gap is $\sim$1-2 eV.  The degeneracy
of each of these levels (t$_{2g}$, e$_g$) is further lifted by distortions such as the ones occurring in 
GdFeO$_3$, or arising through the Jahn-Teller effect or spin-orbit coupling. The energy cost for such
distortion induced splitting is a few meV. Recently, Pavarini et al.\cite{pavarini2} studied crystal field effects in d$^1$ type perovskites such as SrVO$_3$, CaVO$_3$, LaTiO$_3$ and YTiO$_3$. 
It was found that crystal field effects and cation-covalency (GdFeO$_3$ -type distortion) lift the
orbital degeneracy and reduce the orbital fluctuations. Thus, investigating crystal field effects in
model Hamiltonians\cite{PhysRevB.78.045115,PhysRevB.88.195116} is highly relevant for 
understanding of real materials.

\begin{figure}[htbp]
\centering 
\includegraphics[width=\columnwidth]{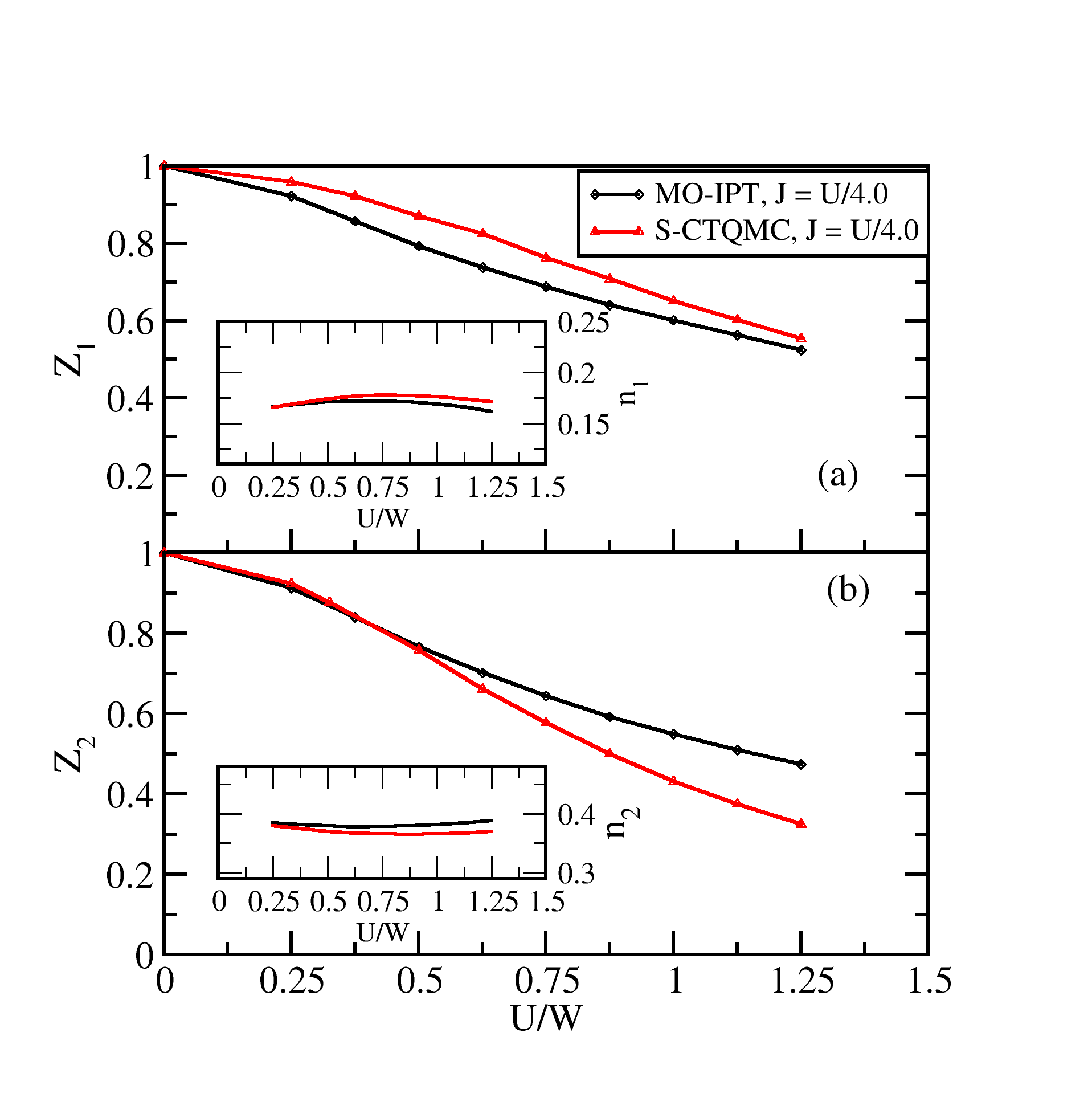}
\caption{(color online) Crystal field effects. Quasi particle weights for (a) orbital-1 and (b) orbital-2, obtained
from MO-IPT and CTQMC for various $U/W$ values with $J=U/4$ at $\beta$=64. The insets show the
corresponding occupancies.} 
\label{fig:fig30} 
\end{figure}

We have investigated the Hamiltonian in Eq.~(\ref{eq:2ohm}) by considering two orbitals with
energies $\epsilon_1 = 0.0$ and $\epsilon_2=-0.2W$, which corresponds to a crystal field splitting
of $0.2W$.  The results from MO-IPT, for a fixed total filling of $n_{tot}=1.1$, are compared with
those from strong coupling CTQMC at the corresponding orbital occupancies. In Fig.~\ref{fig:fig30}, we compare
the quasi particle weights of the two orbitals obtained from MO-IPT with that of CTQMC. We observe
a better agreement of $Z$ for orbital-1 than for orbital-2. This must be expected, since orbital-1
is further away from particle-hole symmetry than orbital-2. The corresponding orbital occupancies as
a function of increasing interaction (and hence $J$) are shown in Fig.~\ref{fig:fig30}. The
deviation between results from the two methods increases with increasing $U$ and $J (=U/4)$, which
indicates that MO-IPT is almost exact for $U/W\lesssim 0.5$.
\begin{figure}[htbp]
\centering 
\includegraphics[width=\columnwidth]{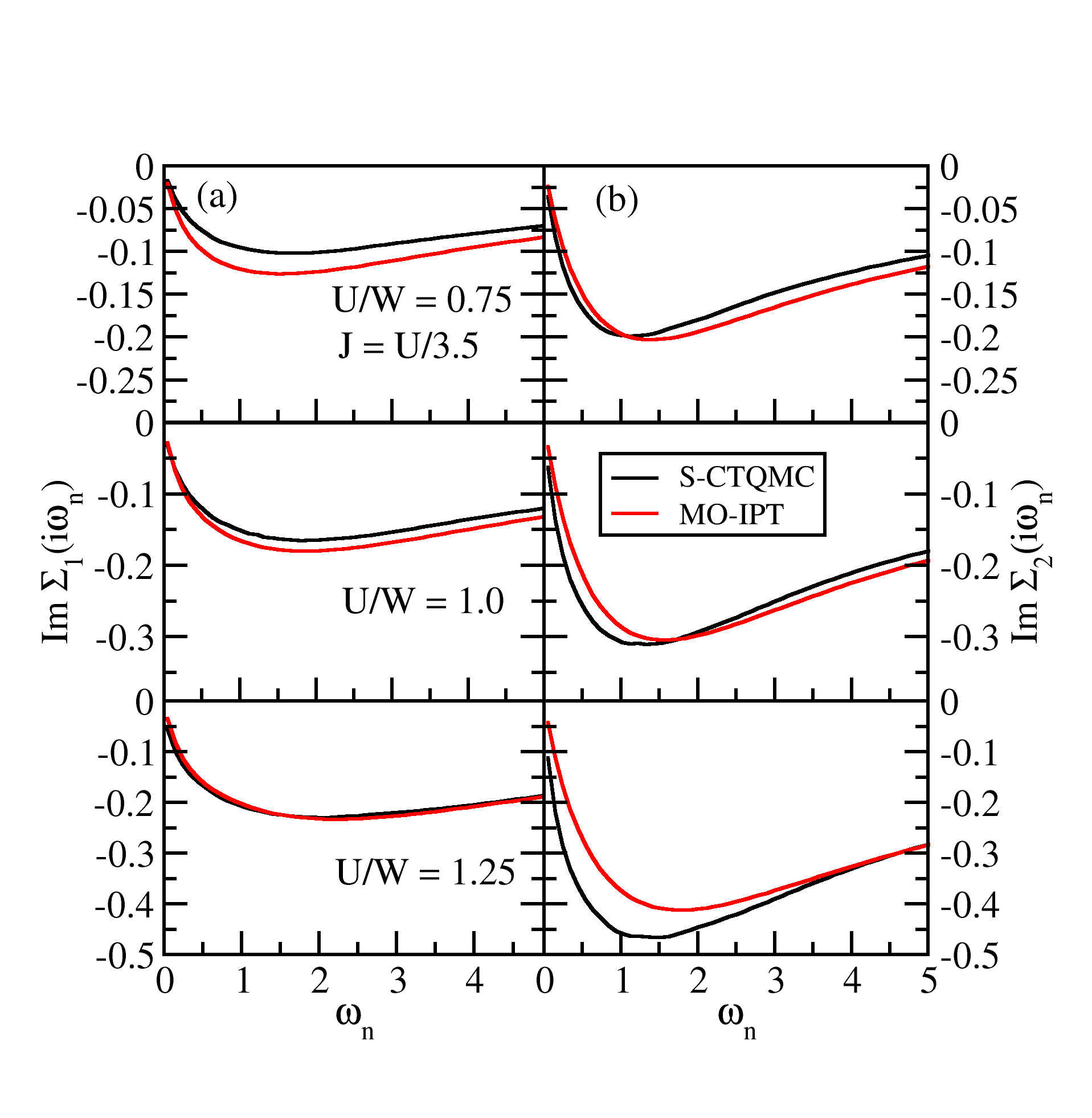} 
\caption{(color online) Crystal field effects. Comparison of the imaginary part of the self
energy for orbital-1 (left) and orbital-2 (right) obtained from MO-IPT and S-CTQMC for various values of $U/W$ and $J = U/3.5$.} 
\label{fig:fig33}
\end{figure}

Next, we benchmark the single-particle dynamics in the presence of crystal field splitting.  In
Fig.~\ref{fig:fig33}, we show the imaginary part of the Matsubara frequency self energies
obtained from MO-IPT and CTQMC for orbitals-1 and 2 (left and right panels respectively). The
agreement between the results is quite evident, this suggesting that the MO-IPT should serve as a
good method to study interacting, real material systems with finite crystal field effects and Hund's
couplings. This is especially true if the material in question has a large number of bands, which
would make it prohibitively expensive to treat with CTQMC, while MO-IPT would be able to handle it
with ease. We now demonstrate the efficacy of MO-IPT when applied to a well studied, real material
system, namely SrVO$_3$. 

\subsection{Application to real materials: SrVO$_3$}
 Over the past decade or so, the combination of density functional theory (DFT) with dynamical mean
field theory, such as LDA+DMFT\cite{Held}, has emerged as one of the most powerful methods for electronic
structure calculations of strongly correlated electronic systems. Although the DFT results contain rich,
material specific information, being a single particle theory, it works well only for weakly
correlated systems where the ratio of Coulomb interaction ($U$) to bandwidth ($W$) is small i.e., 
$U/W \ll 1$. If we consider the opposite limit of $U/W\gg 1$, we have successful methods like the
Hubbard-I and Hubbard-III approximations or the LDA+U method for predicting the ground state of the
system. But these also have limitations, such as the neglect of dynamical fluctuations in the LDA+U
method. In nature, there are many materials, for example transition metal oxides, which lie in
between these two limits. It has been established in the context of model Hamiltonians that the DMFT
can handle both limits quite efficiently. Hence a natural combination of LDA with DMFT is
expected to bring predictive capabilities in the theory of strongly correlated electronic systems.
Nevertheless, LDA+DMFT is not without its own bottlenecks.

 One of the central issues of the LDA+DMFT method is the correct definition of a correlated subspace. The
basic idea of a correlated subspace is to make an appropriate choice of energy window around the
Fermi level and fit the band structure to a few-orbital tight-binding model. Many techniques have
been proposed to construct such a material specific `non-interacting' Hamiltonian. The two major
techniques for this purpose are down-folding \cite{Andersen19951573} and projection based Wannier
function technique \cite{PhysRevB.56.12847}. In general, 
bands which are crossing the Fermi level are considered in the
desired energy window for Hamiltonian construction. For example in transition metal compounds bands having 
d-orbital character normally cross the Fermi level. 
This process becomes simple if there is no
hybridization in the system and these bands with d-orbital character are well
separated from other bands like the p-bands. As Dang et al.\cite{PhysRevB.90.125114}
pointed out, a mixing of these d orbital bands with p orbital bands can
create several complications. 

After getting the 'non-interacting' Hamiltonian , one can add
various types of interactions terms to obtain a full material-specific
multi-orbital model. The solution of such a Hamiltonian is however a major challenge and this
is where the MO-IPT can be most useful, since it scales only algebraically with increasing number of
bands, while yielding real frequency quantities directly. In contrast, impurity solvers like CTQMC
and ED scale exponentially with increasing number of orbitals and are  very expensive,
especially for investigations of real materials. As a test case, we study SrVO$_3$ which is
considered a prototypical example of a strongly correlated electronic system.

\subsubsection*{\bf{(a). Computational Details}}
 We perform our density functional theory (DFT) calculations with linearized augmented plane wave
(LAPW) based method as implemented in the all-electron package WIEN2K\cite{Blaha}. The
experimentally determined structure\cite{Rey1990101} of cubic SrVO$_3$ in a non-magnetic phase 
was used for the calculations (neglecting spin-orbit coupling).
The product of the plane-wave cut off $(K_{max})$ and the smallest atomic sphere radius $(R_{MT})$ was
chosen as $R_{MT}\times K_{max}=7.0$ for controlling the basis set. The radii of the muffin-tin
spheres were chosen to be $10-15\%$ larger than the corresponding atomic radii. Thus, the values
used for $R_{MT}$ were $2.50$ for Sr, $1.89$ for V and $1.71$ for O. With these parameters, charge
leakage was absent and our DFT results agree well with DFT calculation using other basis
sets \cite{0953-8984-20-13-135227}. We utilize the generalized
gradient approximation (GGA) of Perdew, Burke and Ernzerhof\cite{perdew} for the exchange and
correlation functional. In this calculation, we consider 512 {\bf{k}}-points in the irreducible part
of the Brillouin zone. After getting the Bloch-eigen states, all the necessary inputs for
constructing the maximally localized Wannier functions (MLWFs) are prepared by the WIEN2WANNIER
code\cite{kune1}. Finally, the Hamiltonian ${\cal H_{DFT}}$ is constructed in the maximally
localized Wannier basis by taking a projection of the three $V-t_{2g}$ orbitals within the energy window
of -1.0 eV to 1.8 eV  with the standard procedure implemented in
Wannier90\cite{Mosti}. We begin by discussing the DFT results.

\subsubsection*{\bf{(b). GGA+DMFT: Results and discussion}}
\begin{figure}[htbp]
\centering 
\includegraphics[width=\columnwidth]{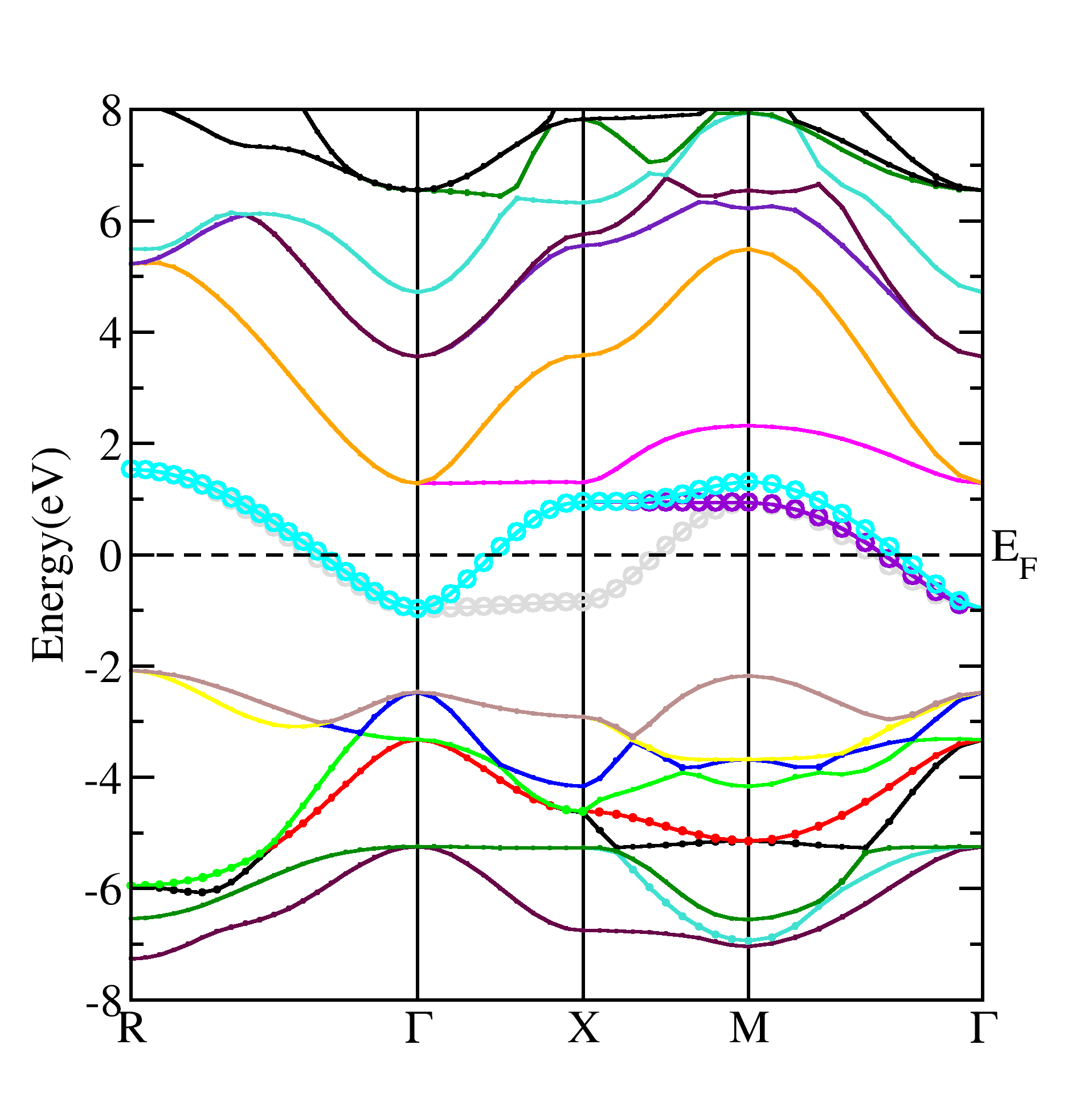} 
\caption{(color online) Band structure of SrVO$_3$ obtained from DFT.} 
\label{fig:fig37} 
\end{figure}
\begin{figure}[htbp]
\centering 
\includegraphics[width=\columnwidth]{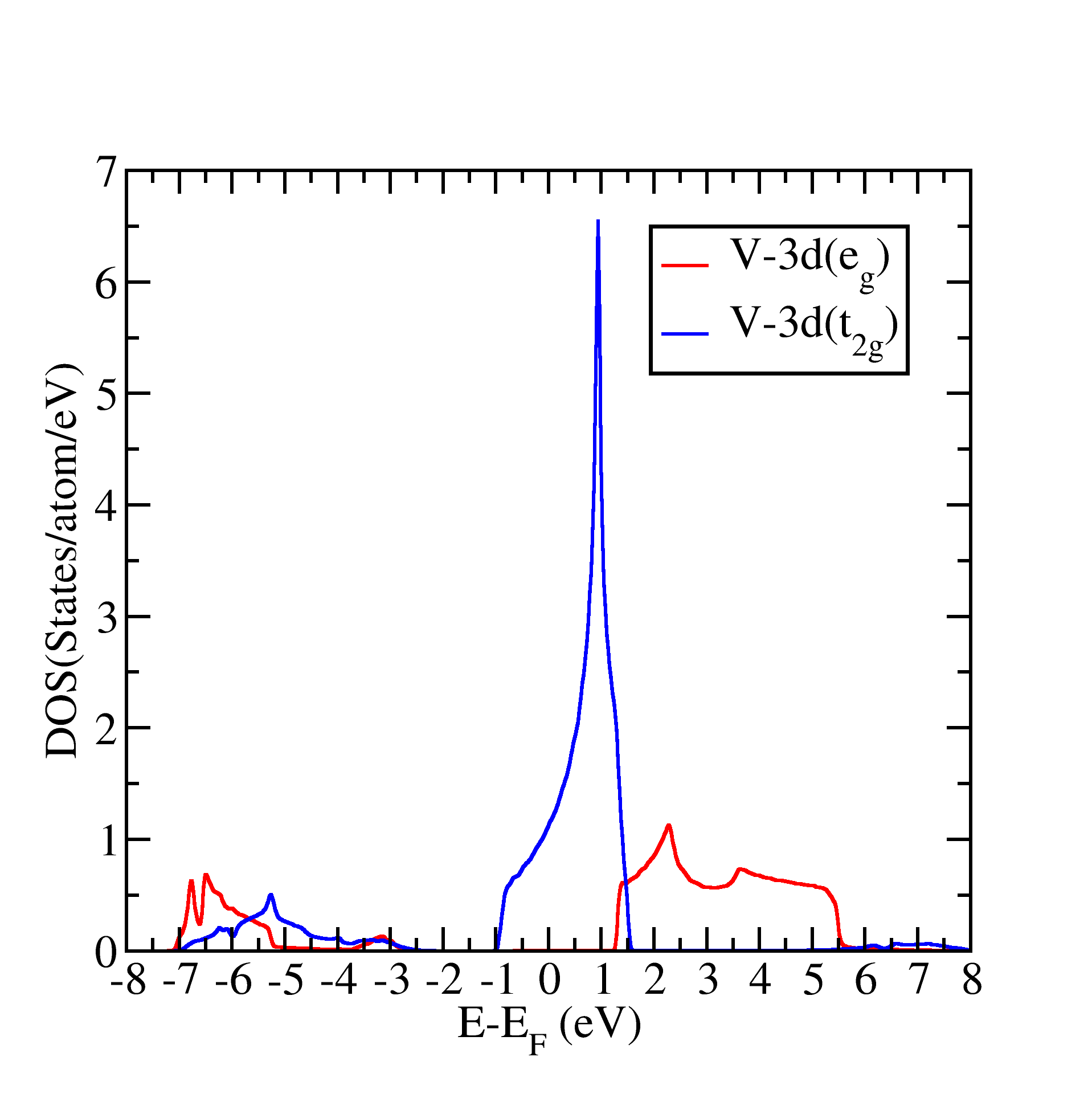} 
\caption{(color online) The projected density of states (DOS) of SrVO$_3$ as calculated by GGA (LAPW).}
\label{fig:fig38} 
\end{figure}

Our computed band structure and density of states (DOS) are presented in Fig.~\ref{fig:fig37}
and Fig.~\ref{fig:fig38}. The three bands, crossing the Fermi level, are highlighted in cyan,
violet and grey colors. These bands originate from the $V-t_{2g}$ states, and are located between
-1.1eV and 1.5eV. The $V-e_g$ states lie at higher energies, between 1.1eV to 5.8eV (see the
projected density of states in Fig.~\ref{fig:fig38}). The band structure agrees well with previous
results by Ishida et. al.\cite{PhysRevB.73.245421} obtained in the LAPW basis. When compared with
results from the linear muffin-tin orbital (LMTO) calculations of Nekrasov et al.
\cite{PhysRevB.72.155106}, the position of $V-t_{2g}$ bands agrees well but the position of $V-e_g$
states differs by about 0.3 eV. This discrepancy is, most likely, due to the difference in basis
sets used in the two calculations. A significant computational simplification results from ignoring the 
hybridization between $V-t_{2g}$ and $V-e_g$ orbitals, since the low energy correlated
subspace comprises just the three $V-t_{2g}$ orbitals.
\begin{figure}[htbp]
\centering
\includegraphics[width=\columnwidth]{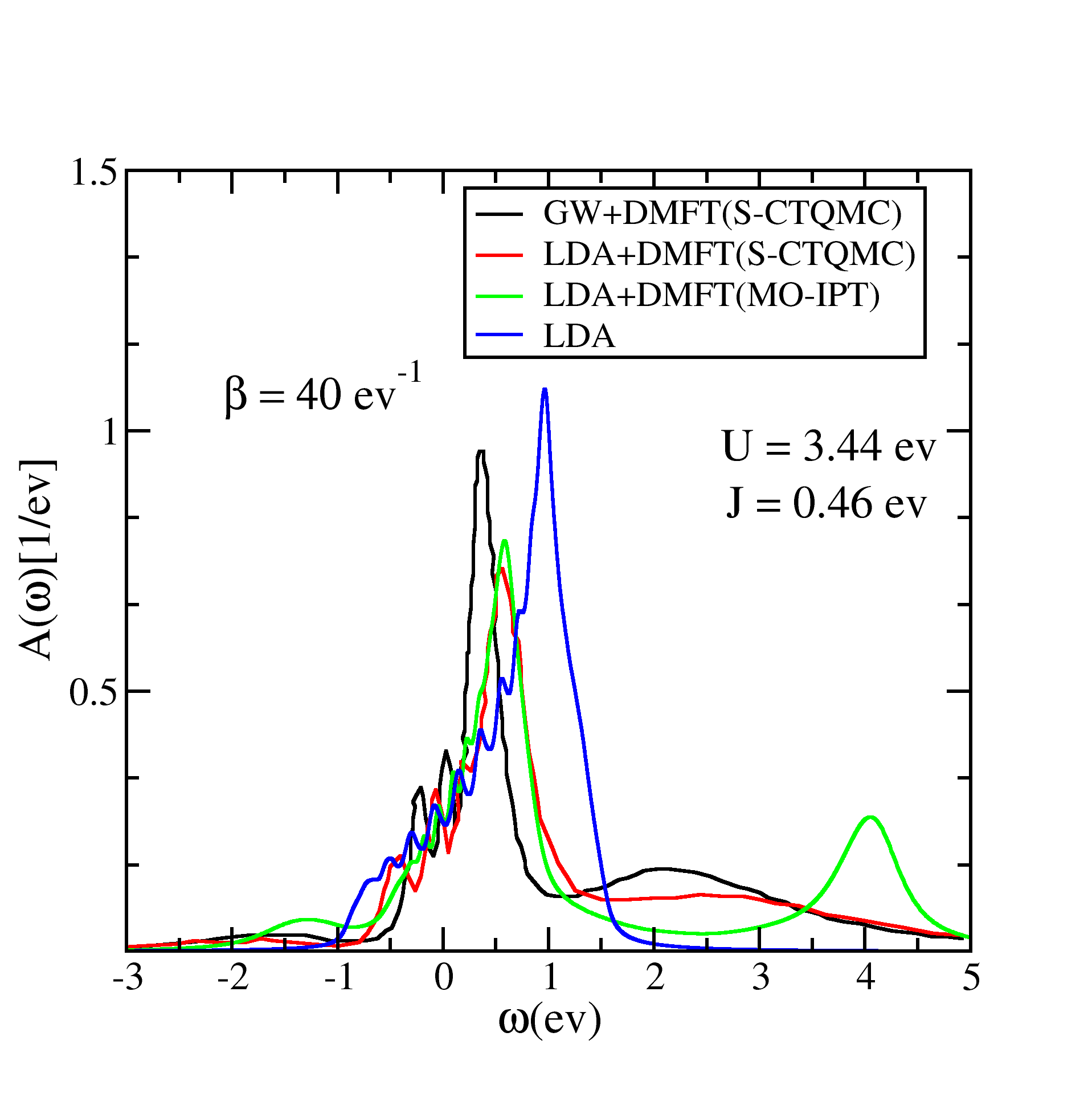}
\caption{(color online) Comparison of spectral function of SrVO$_3$ obtained from different methods for U = 3.44 eV and J = 0.46 eV (see text for details).} 
\label{fig:fig39} 
\end{figure}
Thus, the DFT results yield a `non-interacting' Hamiltonian
${{\hat{H}}}_{DFT}({\mathbf{k}})$, which in this case is a $3\times 3$ matrix for each ${\bf{k}}$.
 Thus, the full $DFT+DMFT$ Hamiltonian is given by
\begin{equation}
{\hat{\cal H}}={{\hat{H}}}_{DFT}({\bf{k}})+{{\hat{H}}}_{int}\,, 
\label{eq:dftham} 
\end{equation}
where ${{\hat{H}}}_{int}$ is the interaction term is given by 
\begin{equation}
{{\hat{H}}}_{int}= U\sum_{i,\alpha}n_{i\alpha\uparrow}n_{i\alpha\downarrow}+\sum_{i\alpha\neq\beta,\sigma
\sigma^\prime}(U^\prime-\delta_{\sigma\sigma^\prime}J)n_{i\alpha\sigma}n_{i\beta\sigma^\prime}\,.
\end{equation}
In the above expression, $i$ stands for V sites and $\alpha$ is the $t_{2g}$ orbital index with spin
$\sigma$. $U$, $U^\prime( = U-2J)$ and  $U^\prime-J( = U-3J)$ are the local, intra orbital and inter orbital Coulomb repulsion
respectively and $J$ is the Hund's exchange.  
The local, non-interacting lattice Green's function, in the orbital basis,
$({\hat{G}}_0(\omega))$, can be obtained from the DFT calculated
${\hat{H}}_{DFT}(\mathbf{k})$ by the following equation as 
\begin{align}
{\hat{G}}_0(\omega)&=\nonumber\\&\sum_{\mathbf{k}}\left(\left[(\omega^+ + \mu)\mathbb{I}
-{{\hat{H}}}_{DFT}({\mathbf{k}})-{{\hat{H}}}_{DC}\right]^{-1}\right)\\
 \equiv &
\left[(\omega^+ + \mu)\mathbb{I} - {\hat{\Delta}}(\omega)\right]^{-1}\,,  
\end{align} 
where $\mu$
is the chemical potential and ${\hat{\Delta}}(\omega)$ is the hybridization. In the DFT approach 
electronic correlations are partially entered through the LDA/GGA exchange-correlation potential. This part of the 
interaction ($\hat{H}$$_{DC}$) has to be subtracted in the LDA+DMFT approach to avoid double-counting. This is not 
an important issue when the low energy effective Hamiltonian contains only the d-manifold because we can absorb it 
into the chemical potential. However it is an important issue when the low energy effective Hamiltonian contains 
O-2p orbitals also. Various schemes for finding the double-counting correction $\hat{H}$$_{DC}$ exist, each with a 
different physical motivation. Details about such schemes may be found in the work by Lechermann et 
al.\cite{PhysRevB.74.125120} and Nicolaus Parragh\cite{PhysRevB.86.155158,Parragh}. In general we can construct the modified host 
Green's function for the $\alpha^{\rm th}$ orbital as 
\begin{equation}
{\tilde{\mathcal{G}}}_{\alpha}=\left(\left[{\hat{G}}^{-1}_{0}+\hat{\epsilon}+{{\hat{H}}}_{DC}-\left(\mu-\mu_0\right)\mathbb{I}\right]^{-1}\right)_{\alpha \alpha}\,.
\end{equation} 
We find the pseudo-chemical potential using the same procedure as in the model calculations. 
  The self-energy can be found, e.g. through the MO-IPT method outlined in Section~\ref{model}.
The second-order self-energy $\Sigma^{(2)}_{\alpha\beta}$ in Eq.~(\ref{eq:Ansat}) is a functional
of the modified host Green's functions, $\left\{{\bf{\tilde{\cal{G}}}}_\alpha\right\}$.
The full local Green's function for the lattice Hamiltonian
(Eq.~(\ref{eq:dftham})) is given by 
\begin{equation}
\hat{G}=\sum_{\mathbf{k}}\left[(\omega^+ + \mu)\mathbb{I}
-{{\hat{H}}}_{DFT}({\mathbf{k}}) - {{\hat{H}}}_{DC}-
{{\hat{\Sigma}}}(\omega)\right]^{-1}\,. 
\label{eq:locG} 
\end{equation}
The above Green's function may be used to obtain a new host Green's function through the Dyson's
equation:
\begin{equation} 
{{{\tilde{\cal{G}}}}}(\omega) =
\left[{{\hat{G}}}^{-1}+{\hat{\Sigma}}+{\hat{H}}_{DC}+\hat{\epsilon}-\left(\mu-\mu_0\right)\mathbb{I}\right]^{-1}\,.
\label{eq:dyson} 
\end{equation}
In general, the chemical potential, $\mu$ is found by fixing the total occupancy from the full
Green's function, ${\hat{G}}$ to be equal to the value found from DFT, 
\begin{equation}
-\frac{1}{\pi}{\rm Im}\int^0_{-\infty} {\rm Tr}{\hat{G}}= n_{tot}^{DFT}\,, 
\label{eq:occonst}
\end{equation} 
where the trace is over spin and orbital indices.

Thus the full solution of the problem proceeds as follows. Given the ${\hat{H}}_{DFT}({\bf
k})$, we guess an initial self-energy, as well as the $\mu$ and $\mu_0$; and use these to find the
local and the host Green's functions through Eqs.~(\ref{eq:locG}) and (\ref{eq:dyson}). The
host Green's functions are then used to find the self-energy, ${\hat{\Sigma}}$ and
Eqs.(\ref{eq:locG}) and (\ref{eq:occonst}) are used to find the chemical potential. For a
fixed $\mu_0$, these equations are then iterated, until the self-energy converges. With the chosen
pseudo-chemical potential, the Luttinger's integral, Eq.~(\ref{eq:lutt}) is computed using the
converged self-energy and local Green's functions.  If the Luttinger's theorem is satisfied within a
numerical tolerance, the solution is considered to be obtained, else the $\mu_0$ is tuned, and the
DMFT equations are iterated, until the Luttinger's theorem is satisfied.
 
 The DFT predicted occupancy per spin on the three correlated V-t$_{2g}$ orbitals in SrVO$_3$ is 0.166, 
which implies SrVO$_3$ is a d$^1$ system. For the DMFT calculations, we employ interaction parameters $U=3.44$ eV and $J=0.46$ eV,
that were obtained by Taranto et al. \cite{Taranto} through the random phase approximation (RPA). For
SrVO$_3$, we have not introduced explicit double counting correction because we choose the correlated
subspace that is identical with the set of Wannier bands. We absorb the double counting correction and
orbital energies in the lattice chemical potential, which we find by using Eq.~(\ref{eq:occonst}).

Our computed GGA+DMFT spectrum for SrVO$_3$ is shown in Fig.~\ref{fig:fig39} and compared with
results obtained from other impurity solvers. The GGA result (shown in blue) has no signatures of
correlation, while each of the DMFT calculations exhibit a three peak structure. The CTQMC results
from GW+DMFT (black) agree qualitatively with those from LDA+DMFT. However, the details do differ.
Namely, the positions and weights of the resonance at the Fermi level and of the Hubbard bands
differ to a significant extent. This difference, naturally, can be attributed to the different
starting points, namely GW {{\it} vs} LDA, of the CTQMC calculations.   Results from the MO-IPT solver
agree with those from CTQMC in the neighborhood of the chemical potential as well as in the
proximity of the lower Hubbard band. The upper Hubbard band is clearly in disagreement with the
CTQMC results. 

\begin{figure}[htbp]
\centering 
\includegraphics[width=\columnwidth]{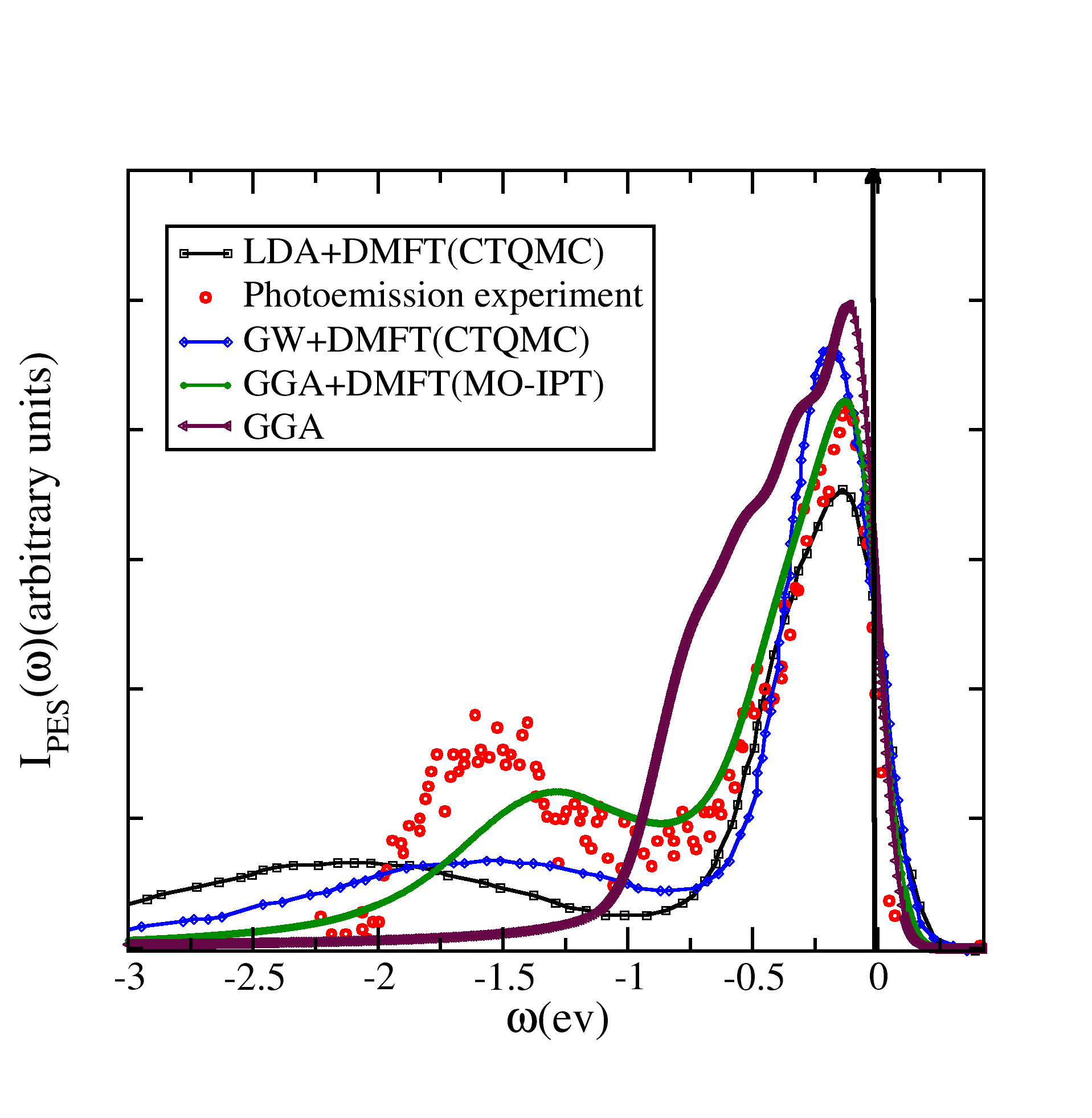} 
\caption{(color online) Comparison of photo emission spectra obtained from different methods GW+DMFT
\cite{Taranto}, GGA+DMFT(MO-IPT), LDA+DMFT(CTQMC)\cite{Taranto} and experiment
\cite{PhysRevLett.93.156402}.}
\label{fig:fig41} 
\end{figure}

As a final benchmark of the GGA+DMFT(MO-IPT) calculation, we compare our result with the
experimentally measured photo emission spectrum (PES) which is shown in Fig.~\ref{fig:fig41}.  A
Hubbard satellite at $\sim-1.5$ eV is seen in the experimental PES spectrum. Our GGA+DMFT(MO-IPT)
calculation predicts the Hubbard satellite at -1.25 eV. Results from other approaches, namely LDA,
LDA+DMFT(CTQMC) and GW+DMFT(CTQMC) are also reproduced.  Surprisingly, the closest match with the
experiment is achieved by the GGA+DMFT(MO-IPT) in terms of the position and width of the resonance
at the Fermi level and of the lower Hubbard band. Thus, we infer that the MO-IPT method outlined in
this work may be used as an efficient tool to study the electronic structure of real material
systems.

\section{Conclusions} 

The development of iterated perturbation theory as an impurity solver for
single band models and for multi-band models dates back to almost two decades. Although a few
comparisons with numerically exact methods have been made, being a perturbative approach, the method
has suffered from reliability issues, especially for multi-orbital systems. Nevertheless,  several
multi-orbital extensions of IPT have been proposed and used to investigate model Hamiltonians and
even real material systems.  In this work, we have outlined a multi-orbital extension of IPT, and 
benchmarked it extensively against continuous time quantum Monte Carlo results.  Our work 
is the first systematic study of the multi-orbital Hubbard model using MO-IPT as a solver and 
varying parameters such as filling, Hund's coupling, and Coulomb repulsion, as well as including crystal 
field effects and application to real materials. One of the main
bottlenecks in methods based on spectral moment expansions is the evaluation of high-order
correlation functions. We find that including such correlations that are beyond two-particle type
through approximate methods such as CPA or lower order decomposition, can lead to spurious features
at the chemical potential. We find the best benchmarks simply by neglecting correlations beyond
two-particle.  We conjecture that evaluation of the higher-order correlations through exact methods
such as ligand field theory might be able to circumvent the issues mentioned above\cite{PhysRevB.77.195124,PhysRevB.57.6884}.  
We are presently implementing such a procedure. This procedure will also enable us to treat the Hund's coupling
term in the rotationally invariant form rather than the simpler and approximate density-density type
treated in the present work. We are also planning to extend our method to incorporate 
the off-diagonal hybridization, which  at present is very difficult to handle by numerical exact methods 
like CTQMC. Apart from the benchmarks for model Hamiltonians in various parameter
regimes, we have also carried out a GGA+DMFT(MO-IPT) study of the perovskite SrVO$_3$, and compared
our predicted photoemission spectrum with experiments and results from other methods. 
The agreement with experiments was
found to be excellent. A full scale implementation of the method outlined here, with detailed
instructions for installation and use may be found at
http://www.institute.loni.org/lasigma/package/mo-ipt/.

\section{Acknowledgments} 

We thank CSIR and DST (India) for research funding. Our simulations
used an open source implementation\cite{Hafer} of the hybridization expansion continuous-time
quantum Monte Carlo algorithm\cite{Comanac} and the ALPS \cite{Bauer} libraries. 
This work is supported by NSF DMR-1237565 and NSF EPSCoR Cooperative Agreement EPS-1003897. 
Supercomputer support is  provided by the Louisiana Optical Network Initiative (LONI) and HPC@LSU.
NSV acknowledges an international travel fellowship award from IUSSTF. 
Dasari acknowledges the hospitality of the department of Physics \& Astronomy 
and the Center for Computation $\&$ Technology, 
at Louisiana State University. 

\section{Appendix}
In this appendix, we provide the derivations of the unknown parameters appearing in the MO-IPT ansatz for the self energy (Eq.~(\ref{eq:Ansat})).  

\noindent \underline{\textbf{Derivation for A$_\alpha$:}}\\

 The spectral representation of the $\alpha^{\rm th}$-orbital Green's function is given by
\begin{equation} G^{\alpha
\alpha}(z)=\int^\infty_{-\infty}\frac{D^{\alpha \alpha}(\epsilon)d\epsilon}{z-\epsilon}\,.
\label{eq:Hilbert} 
\end{equation}
This can be Taylor expanded to obtain the Green function in terms
of spectral moments,
\begin{equation} 
G^{\alpha \alpha}(z)=\int^\infty_{-\infty}\frac{D^{\alpha
\alpha}(\epsilon)d\epsilon}{z}(1+\frac{\epsilon}{z}+\frac{\epsilon^2}{z^2}+\cdots)=\sum^\infty_{n=0}\frac{\mu_n}{z^{n+1}}\,,
\label{eq:momexp} 
\end{equation}
where $\mu_n$'s are the spectral moments.  We can also represent
the Green function in terms of a continued fraction expansion and this is given by
\begin{equation}
G^{\alpha \alpha}(z)=\frac{\alpha_{1}}{z+\frac{\alpha_{2}}{1+\frac{a_3}{z+a_4+\cdots}}}\,.
\label{eq:gfncf} 
\end{equation}
By comparing Eq.~(\ref{eq:gfncf}) with Eq.~(\ref{eq:momexp}), we obtain the continued fraction expansion 
coefficients in terms of
spectral moments. Now we can calculate the spectral moments exactly up to any order by using the
following expressions\cite{Nolting1}:
\begin{equation*} 
\mu^{\alpha \alpha}_n=
\left\langle\left[\underbrace{{[...[[f_{\alpha},{\cal{H}}],{\cal{H}}],...{\cal{H}}]}_-}_\text{(n-p)-fold},{\underbrace{{[{\cal{H}},...[{\cal{
H}},[{\cal{ H}},{f^{\dagger}_{\alpha}}]]...]}_-}_\text{p-fold}}\right]_+\right\rangle
\end{equation*}
$\centerline{n = 0,1,2,.... \hspace{0.5mm}; \hspace{0.5mm} 0$\leq$ p $\leq$ n}$ The
relation between the first few spectral moments and the continued fraction expansion coefficients is given
by,
\begin{equation}
\alpha_1=\mu^{\alpha \alpha}_{0}=\langle \{f_{\alpha}, f^{\dag}_{\alpha}\}\rangle = 1\,, 
\label{eq:momone} 
\end{equation}
\begin{equation*}
\alpha_2=-\mu^{\alpha \alpha}_{1}=\langle \{[f_{\alpha},{\cal H}_{imp}], f^{\dag}_{\alpha}\}\rangle\,,
\end{equation*}
\begin{equation} 
\alpha_2=-[(\epsilon_{\alpha}-\mu)+\sum_{\beta\neq(\alpha)} U_{\alpha \beta} \langle n_{\beta}\rangle]\,, 
\label{eq:momtwo} 
\end{equation}
\begin{equation}
\alpha_3=-\frac{\mu^{\alpha \alpha}_2\mu^{\alpha \alpha}_0-(\mu^{\alpha \alpha}_1)^2}{\mu^{\alpha \alpha}_1\mu^{\alpha \alpha}_0}\,, 
\label{eq:momthree} 
\end{equation}
\begin{align} 
\mu^{\alpha\alpha}_{2}= (\epsilon_{\alpha}-\mu)^2 &+\frac{1}{N}\sum_{k\alpha}
V^{2}_{k\alpha}+2(\epsilon_{\alpha}-\mu) \sum_{\beta\neq \alpha} U_{\alpha\beta} \langle n_{\beta
}\rangle\nonumber\\&+\sum_{\beta\neq \alpha}\sum_{\gamma\neq \alpha} U_{\alpha \beta} U_{\alpha
\gamma}\langle n_{\beta}n_{\gamma}\rangle\,,
\label{eq:momtwoexp} 
\end{align}
\begin{equation} 
\alpha_3=\frac{\frac{-1}{N}\sum_{k\alpha}
{V^{2}}_{k\alpha}-{\sum}_{\beta,\gamma\neq \alpha}U_{\alpha \beta} U_{\alpha \gamma} \left(\langle
n_{\beta}n_{\gamma}\rangle-\langle n_{\beta}\rangle \langle
n_{\gamma}\rangle\right)}{(\epsilon_{\alpha}-\mu)+\sum_{\beta\neq\alpha} U_{\alpha \beta} \langle
n_{\beta}\rangle}\,. 
\label{eq:momthreeexp} 
\end{equation}
For sufficiently large values of $z$, one
can truncate the continued fraction expansion of the Green's function (Eq.~(\ref{eq:gfncf})) at
the appropriate level and take the limit $z\rightarrow$ $\infty$. Up to the second order moment

\begin{equation} G^{\alpha
\alpha}(z)=\frac{\alpha_{1}}{z+\alpha_{2}-\frac{\alpha_2\alpha_3}{z}}. 
\label{eq:gfcnf_trunc}
\end{equation}
After substituting the continued fraction expansion coefficients in
Eq.~(\ref{eq:gfcnf_trunc}), we find the self energy contribution to the Green's function in the
high frequency limit as
\begin{align*}
\Sigma_{\alpha}(\omega)\xrightarrow{\omega\rightarrow\infty}&\sum_{\beta\neq \alpha} U_{\alpha
\beta} \langle n_{\beta}\rangle\nonumber\\&+\frac{\sum_{\beta,\gamma\neq\alpha}U_{\alpha \beta}
U_{\alpha \gamma} \left(\langle n_{\beta }n_{\gamma}\rangle-\langle n_{\beta}\rangle \langle
n_{\gamma}\rangle\right)}{\omega},
\end {align*}
\begin{align}
\Sigma_{\alpha}(\omega)=&\sum_{\beta\neq\alpha} U_{\alpha \beta} \langle
n_{\beta}\rangle+\frac{\sum_{\beta\neq\alpha} U_{\alpha \beta}^2 \langle n_{\beta}\rangle\left(1-
\langle
n_{\beta}\rangle\right)}{\omega}\nonumber\\&+\frac{\sum_{\beta\neq\alpha}\sum_{\gamma\neq(\beta\neq\alpha)}
U_{\alpha \beta} U_{\alpha \gamma}\left(\langle n_{\beta}n_{\gamma}\rangle-\langle n_{\beta}\rangle
\langle n_{\gamma}\rangle\right)}{\omega}.
\label{eq:sigmahighfre} 
\end{align}
In the high frequency limit the self energy ansatz reduces to the following form:
\begin{equation} 
\Sigma_{\alpha}(\omega)=\sum_{\beta\neq(\alpha)} U_{\alpha \beta} \langle
n_{\beta}\rangle+A_{\alpha}\sum_{\beta\neq(\alpha)}\Sigma^{(2)}_{\alpha \beta}\,. 
\label{eq:highfeq}
\end{equation}
It is easy to show that in the limit of high frequencies, $\Sigma^{(2)}_{\alpha \beta}$ has the following
form\cite{Martin3},
\begin{equation}
\Sigma^{(2)}_{\alpha \beta}= \frac{U_{\alpha \beta}^2}{\omega}\langle
n_{0\beta}\rangle\left(1-\langle n_{0\beta}\rangle\right). 
\label{eq:pairbubble} 
\end{equation}
Here n$_{0\beta}$ is the Hartree-corrected impurity charge 
because the propagators used in the second order pair bubble diagram are Hartree-corrected
propagators. 
We obtain the expression for $A_\alpha$ by substituting Eq.~(\ref{eq:pairbubble}) in Eq.~(\ref{eq:highfeq}) 
and comparing with Eq.~(\ref{eq:sigmahighfre}) as
\begin{align*}
A_{\alpha}= &\frac{\sum_{\beta\neq(\alpha)}U_{\alpha \beta}^2 \langle
n_{\beta}\rangle\left(1-\langle n_{\beta} \rangle \right)}{\sum_{\beta\neq(\alpha)} U_{\alpha
\beta}^2 \langle n_{0\beta} \rangle \left(1-\langle n_{0\beta} \rangle
\right)}\nonumber\\&+\frac{\sum_{\beta\neq(\alpha)} U_{\alpha
\beta}\sum_{\gamma\neq(\beta\neq\alpha)} U_{\alpha \gamma}\left(\langle
n_{\beta}n_{\gamma}\rangle-\langle n_{\beta}\rangle\langle
n_{\gamma}\rangle\right)}{\sum_{\beta\neq(\alpha)} U_{\alpha \beta}^2 \langle n_{0\beta}\rangle
\left(1-\langle n_{0\beta}\rangle \right)}\,. 
\end{align*}
Note that a two-particle correlation
function is needed to find $A_\alpha$.

\noindent \underline{\textbf{Derivation for B$_{\alpha}$:}}\\
 
The relation between the impurity Green's function and the self energy in the atomic limit is,
\begin{equation} 
G_{\alpha}(\omega)=\frac{1}{\omega^++\mu-\epsilon_{\alpha}-\Sigma_{\alpha}(\omega)}\,,
\label{eq:atomicG} 
\end{equation}
where the self-energy, $\Sigma_{\alpha}(\omega)$ may be
represented as a continued fraction:
\begin{equation}
\Sigma_{\alpha}(\omega)=\omega^++\mu-\epsilon_{\alpha}-\frac{1}{\frac{\alpha_{1}}{z+\frac{\alpha_{2}}{1+\frac{a_3}{z+a_4\cdots}}}}\,.
\label{eq:atomicS}
\end{equation}
As a simple case we consider only two poles in the self energy. In
principle we can keep all the poles of the self energy but the difficulty is that a pole of order $n$
involves the $(n+1)^{\rm th}$ order correlation function. These functions are very
hard to calculate without making approximations. With the two pole approximation for the self energy,
Eq.~(\ref{eq:atomicS}) reduces to the following form\cite{Oudovenko}
\begin{equation}
\Sigma_{\alpha}(\omega)=\sum_{\beta\neq(\alpha)}U_{\alpha \beta} \langle n_\beta \rangle
+\frac{\alpha_2\alpha_3}{\omega^++\alpha_3+\alpha_4} 
\label{eq:S_mom} 
\end{equation}
\begin{align} 
{\rm where}\hspace{1.5mm}\alpha_2 = -\mu^{\alpha \alpha}_1\,, 
\label{eq:alpha2} 
\end{align}
\begin{equation}
\alpha_3=-\frac{\mu^{\alpha \alpha}_2-(\mu^{\alpha \alpha}_1)^2}{\mu^{\alpha \alpha}_1}\,,
\label{eq:alpha3} 
\end{equation}
\begin{align} 
\rm{and} \hspace{1.5mm} \alpha_4=-\frac{\mu^{\alpha\alpha}_1 \mu^{\alpha \alpha}_3-(\mu^{\alpha \alpha}_2)^2}{\mu^{\alpha
\alpha}_1\mu^{\alpha \alpha}_2-(\mu^{\alpha \alpha}_1)^2}\,.
\label{eq:alpha4} 
\end{align}
In the atomic limit (V$\rightarrow$ 0), the second order pair bubble diagram $\Sigma^{(2)}_{\alpha\beta}(\omega)$ reduces to the following
form\cite{Kajueter1,Martin1},
\begin{equation}
\Sigma^{(2)}_{\alpha\beta}(\omega)=\frac{U^2_{\alpha\beta}[\langle n_{0\beta}\rangle(1-\langle
n_{0\beta}\rangle )]}{\omega^+ + \mu_0}\,. 
\label{eq:atomic_pair} 
\end{equation}
Here $\mu_0$ is the
pseudo-chemical potential. As mentioned earlier, we find this quantity by satisfying the Luttinger's
theorem or equivalently the Friedel's  sum rule. Now, the self energy ansatz becomes
\begin{eqnarray} 
\Sigma_\alpha&=&\sum_{\beta\neq(\alpha)}U_{\alpha \beta} \langle n_\beta \rangle  + \nnu \\
&&\frac{A_\alpha \sum_{\beta\neq(\alpha)}U^2_{\alpha\beta}[\langle n_{0\beta}\rangle (1-\langle
n_{0\beta}\rangle )]}{\omega^++\mu_0-B_{\alpha}\sum_{\beta\neq (\alpha)}U^2_{\alpha\beta}[\langle
n_{0\beta}\rangle (1-\langle n_{0\beta}\rangle )]}\,.
\label{eq:sigma_alpha} 
\end{eqnarray}
By comparing Eq.~(\ref{eq:sigma_alpha}) with
Eq.~(\ref{eq:atomicS}) we find the expression for $B_\alpha$ in terms of spectral moments as,
\begin{equation} 
B_\alpha=\frac{\mu_0-(\alpha_3+\alpha_4)}{\sum_{\beta \neq (\alpha)}U^2_{\alpha\beta}[\langle
n_{0\beta}\rangle (1-\langle n_{0\beta}\rangle )]}.
\label{eq:Balpha} 
\end{equation}
After substituting the spectral moments in Eq.~(\ref{eq:Balpha}) B$_\alpha$ becomes,
\begin{align}
&B_{\alpha}= \frac{\mu_0+\epsilon_{\alpha}-\mu-\sum_{\beta \neq \alpha} U_{\alpha \beta} \langle
n_{\beta} \rangle}{\tau_{\alpha}}\nonumber\\&-\frac{\sum_{\beta \neq \alpha}\sum_{\gamma \neq
\alpha}\sum_{\eta \neq \alpha}U_{\alpha \beta}U_{\alpha \gamma}U_{\alpha \eta} \left[\langle
n_{\beta} \rangle \langle n_{\gamma} n_{\eta} \rangle -\langle n_{\beta} n_{\gamma} n_{\eta}
\rangle\right]}{\tau^2_{\alpha} A_{\alpha}}, 
\label{eq:Balpha_final} 
\end{align}
where
\begin{align} 
\tau_{\alpha}=\sum_{\beta\neq\alpha}U^2_{\alpha \beta} \langle n_{0\beta}\rangle (1-\langle n_{0\beta}\rangle).
\label{eq:tau} 
\end{align}



\bibliographystyle{apsrev4-1}
\bibliography{apssamp} 
\end{document}